\documentclass[epjST]{svjour}
\usepackage{graphics}
\usepackage{amsmath}
\usepackage{amssymb}
\usepackage{hyperref}

\usepackage{color}

\newcommand{\be}{\begin{equation}}
\newcommand{\ee}{\end{equation}}
\newcommand{\bea}{\begin{eqnarray}}
\newcommand{\eea}{\end{eqnarray}}
\newcommand{\rr}{\mathbf{r}}
\newcommand{\kk}{\mathbf{k}}
\newcommand{\KK}{\mathbf{K}}
\newcommand{\apgt}{  \ {\raise-.5ex\hbox{$\buildrel>\over\sim$}}\ }
\newcommand{\aplt}{\ {\raise-.5ex\hbox{$\buildrel<\over\sim$}}\ }

\newcommand{\oo}{\mathbf{0}}

\begin{document}
\title{Spin Squeezing in Finite Temperature Bose-Einstein Condensates : Scaling with the system size}

\author{A. Sinatra\inst{1}\fnmsep\thanks{\email{alice.sinatra@lkb.ens.fr}} \and E. Witkowska\inst{2} 
\and Y. Castin\inst{1} }
\institute{Laboratoire Kastler Brossel, Ecole Normale Sup\'erieure, UPMC and CNRS,  Paris, France \and 
Institute of Physics, Polish Academy of Sciences, Warszawa, Poland}

\abstract{
We perform a multimode treatment of spin squeezing induced by interactions in atomic condensates, and
we show that, at finite temperature, the maximum spin squeezing has a finite limit when the atom number $N\to \infty$ at fixed density and interaction strength. To  calculate the limit of the squeezing parameter for a spatially homogeneous system we perform a double expansion with
two small parameters: $1/N$ in the thermodynamic limit and the non-condensed fraction $\langle N_{\rm nc} \rangle/N$ in the Bogoliubov limit. 
To test our analytical results beyond the Bogoliubov approximation, and to perform numerical experiments, we use improved classical field 
simulations with a carefully chosen cut-off, such that the classical field model gives for the ideal
Bose gas the correct non-condensed fraction in the Bose-condensed regime.
}

\maketitle

\section{Introduction}
A two-level atom can be described as an effective spin $1/2$. Here, to describe an ensemble of atoms in two different internal states $a$ and $b$,
that are typically two hyperfine states, we use the picture of a ``collective spin". This spin, of length $N/2$, is simply the sum of the effective spins $1/2$ that describe the internal degrees of freedom of each atom.
In the second quantized formalism the three hermitian spin components $\hat{S}_x$, $\hat{S}_y$ and $\hat{S}_z$ are defined by:
\begin{eqnarray}
\hat{S}_+\equiv \hat{S}_x+i \hat{S}_y&=& \int d^3r \; {\hat{\psi}}^\dag_a
(\rr\,){\hat{\psi}}_b(\rr\,) \label{eq:SxSy} \\
\hat{S}_z &=&\dfrac{\hat{N}_a-\hat{N}_b}{2}
\label{eq:Sz}
\end{eqnarray}
where the bosonic field operators $\hat{\psi}_{a,b}$ 
obey the usual commutation relations, 
$\hat{N}_a=\int d^3r \, {\hat{\psi}}^\dag_a (\rr\,){\hat{\psi}}_a(\rr\,)$ is the atom number in component $a$ and the same for $b$.
The spin operators are dimensionless and obey the commutation relations $[\hat{S}_x,\hat{S}_y]=i \hat{S}_z$ and cyclic permutations.
Physically $\hat{S}_z$ is the population difference between $a$ and $b$ states, while $\hat{S}_x$ and $\hat{S}_y$ describe one-body coherence between them.

Spin squeezing \cite{Ueda:1993} is about creating quantum correlations, in such an ensemble of atoms, that can be useful for metrology.
In particular spin squeezed states can be used to improve the accuracy of atomic clocks beyond the so called 
``standard quantum limit" that has been already reached in the most precise clocks \cite{Salomon:1999}.
The resulting gain for metrology is quantified 
by a spin squeezing parameter $\xi^2$ \cite{Wineland:1994,Sorensen:2001}:
\begin{equation}
\xi^2=\frac{N \Delta S_{\perp,\mbox{\small min}}^2}{|\langle {\bf S}\, 
\rangle|^2 } \, , \label{eq:xi2def}
\end{equation}
where $N$ is the total atom number and $ \Delta S_{\perp,\mbox{\small min}}^2$ is the minimal variance of the collective spin orthogonally to 
the direction of its mean value $\langle  {\bf S}\,  \rangle$. The state is squeezed if and only if $\xi^2<1$. 
As explained in \cite{Wineland:1994}, in an atomic clock experiment using Ramsey population spectroscopy, $\xi$ directly gives the reduction in 
the statistical fluctuations of the measured frequency $\omega_{ab}$ with respect to using uncorrelated atoms (for the same atom number $N$ and 
the same Ramsey time ${\cal T}$):
\be
\Delta \omega_{ab}^{\rm sq}= \xi \Delta \omega_{ab}^{\rm unc}={\xi \over \sqrt{N} {\cal T} }
\ee
The parameter $\xi$ in Eq.(\ref{eq:xi2def}) is in fact the properly normalized ratio between the ``noise" $\Delta S_{\perp,{\rm min}}$ and the ``signal"  
$|\langle {\bf S} \rangle|$. In experiments $\Delta S_{\perp,{\rm min}}$ is directly measured by measuring $\hat{S}_z$ 
after an appropriate state rotation and $|\langle {\bf S} \rangle|$ is separately deduced from the Ramsey fringes contrast.

Very recently experimental breakthroughs in spin squeezing have been achieved using either the interaction between atoms and
light in an optical cavity \cite{Vuletic:2010} or atomic interactions in bimodal Bose-Einstein condensates  \cite{Oberthaler:2010}, \cite{Treutlein:2010}. 
The ultimate limits of the different paths to spin squeezing are still objects of active studies 
\cite{LiYun:2008,Sinatra:2011,Minguzzi:arxiv,Sinatra:Frontiers,Vuletic:2011}. We address here the issue of non-zero 
temperature and of the influence of the non condensed fraction for spin squeezing schemes using Bose-Einstein condensates.
 
We face the following physical problem: An interacting Bose gas, prepared at finite temperature in the internal state $a$,
is subjected to a sudden $\pi/2$ mixing pulse that puts each atom in a coherent superposition of two different internal states $a$ and $b$.
From this out of equilibrium state, with factorized spin and motional variables, quantum correlations and spin squeezing 
are created dynamically by the atomic interactions \cite{Ueda:1993}, \cite{Sorensen:2001}. Let us first sketch how this happens in a simple 
two-mode picture, i.e. assuming that all the atoms in $a$ or $b$ share the same wave function for their motional degrees of freedom.
After the mixing pulse, the two condensates in $a$ and $b$ have a well defined relative phase, with a relative phase distribution whose width scales as $1/\sqrt{N}$, and fluctuations in the relative particle number difference scaling as $\sqrt{N}$. 
In a two-mode picture, the initial state can be expanded over Fock states $|N_a,N-N_a\rangle$ with $N_a$ particles in state $a$ and $N-N_a$ particles
in state $b$. Due to atom-atom interactions, each Fock state acquires a phase in the evolution that is proportional to $N_a-N_b$ 
\cite{Ueda:1993,CastinDalibard:1997,Sinatra:1998}. This situation 
is completely equivalent to the evolution of a coherent state in a Kerr medium in optics. During the evolution, due to the different phase shifts of the
different Fock states, the relative phase distribution starts to spread. At the same time, quantum fluctuations orthogonal to the mean spin direction get distorted and, before the relative phase distribution has sensibly spread, spin squeezing is created in the sample.    
Our aim is to include the two-mode quantum dynamics that we just described, {\it and} the effect of the thermally excited non-condensed modes {\it within the same formalism}. The thermal modes {\it also} provide a condensate phase
spreading \cite{Kuklov:2000},\cite{Sinatra:2007},\cite{Sinatra:2008},\cite{Sinatra:2009} and are expected to affect the spin squeezing generated in the system at non-zero temperature \cite{Sinatra:2011}. For a review of spin squeezing and decoherence see also \cite{Sinatra:Frontiers}.

A central issue is the {\it scaling of the squeezing} as the system gets large, i.e. in the thermodynamic limit.
Most studies are based on a two-mode description \cite{Ueda:1993}. In this frame the squeezing parameter 
minimized over time $\xi^2_{\rm min}$ tends to zero (infinite metrology gain) for $N\to \infty$ as $\xi^2_{\rm min} \sim N^{-2/3}$.
Although some studies beyond the two-mode theory were performed \cite{Sorensen:2001,Poulsen:2001,Sorensen:2002}  they could not prove or
disprove the two-mode scaling of spin squeezing in real condensates.
Here we can go further. We find that for realistic atom numbers, the two-mode scaling $\xi^2_{\rm min} \sim N^{-2/3}$ is meaningless at finite temperature and that the spin squeezing parameter $\xi^2_{\rm min}$ at the thermodynamic limit has a finite non-zero value that we calculate explicitly. In this paper
we present a detailed derivation of the results given in \cite{Sinatra:2011} and we present new improved classical field simulations,
with a carefully chosen cut-off such that the classical field model gives for the ideal
Bose gas the correct non-condensed fraction in the Bose-condensed regime.
We also present results for the squeezing that would be measured by detecting only the condensed particles,
which we call the ``condensate squeezing",
and we show that it is much worse than the squeezing of the total field for reasons that we explain in the paper. 

In section \ref{sec:problem} we formalize the problem and expose our approach to solve it. In section
 \ref{sec:numexp} we proceed with two numerical experiments. These experiments show (i) the existence of a non-zero thermodynamic limit for 
the squeezing parameter in contrast with the predictions of the two-mode theory, and (ii) the universal scaling with
the temperature of the squeezing in the thermodynamic and weakly interacting limit. Analytical calculations are performed in section \ref{sec:analy}.
By performing a double expansion of $\xi^2$ 
in terms of two small parameters, the inverse atom number $1/N$ controlling the thermodynamic limit
and the non-condensed fraction controlling the weakly interacting limit,
we obtain explicitly the minimal squeezing parameter that it is possible to achieve by this method
as a function of the initial temperature and the interaction strength. A physical interpretation of the results is given in section \ref{sec:interpret}.
In that section we also show that the squeezing defined for the total field and the squeezing defined for the condensate mode only are
very different and we give a physical explanation. We conclude in section \ref{sec:conclusions}.

\section{The problem}
\label{sec:problem}

\subsection{The Quantum Model}
\label{sub:model}
We consider a spatially homogeneous system of $N$ bosons in two internal states that interact with short range binary interactions. 
We take for simplicity identical interactions in components $a$ and $b$ 
and no crossed $a$-$b$ interactions 
\footnote{In ${}^{87}$Rb atoms this may be done by spatial separation of the spin states \cite{Treutlein:2010}
or by Feshbach tuning of the $a$-$b$ scattering length \cite{Oberthaler:2010}.}. 
The system is discretized on a cubic lattice of lattice spacing $l$, with periodic boundary condition of period
$L$ along each direction $x,y,z$. For numerical convenience, $L/l=n_{\rm max}$ is an even integer.
There are in total ${\cal N}\equiv V/dV$ lattice points, where $V=L^3$ is the system volume
and $dV=l^3$ the unit cell volume.
The Hamiltonian for one separate spin component, e.g. component $a$, reads
\begin{equation}
\hat{H}_a=\sum_\kk \frac{\hbar^2 k^2}{2m} \hat{a}^\dagger_\kk \hat{a}_\kk + 
		\frac{g_0}{2} dV \sum_{\rr}  {\hat{\psi}}_a^\dagger(\rr) {\hat{\psi}}_a^\dagger(\rr) \hat{\psi}_a(\rr) \hat{\psi}_a(\rr) \,
\label{eq:discrHam}
\end{equation}
In the kinetic energy term we have expanded the field operator over plane waves
\be
\hat{\psi}_a({\rr})=\sum_\kk \hat{a}_\kk {e^{i\kk \cdot \rr}\over \sqrt{V}} \,
\ee
and $\hat{a}_\kk$ annihilates a particle of wave vector $\kk$ belonging to the first Brillouin zone 
(FBZ) $[-\pi/l,\pi/l[^3$ of the lattice,
so that along each direction $\nu$, $k_\nu$ $\in \frac{2\pi}{L}\{-L/(2l),\ldots$, $L/(2l)-1\}$.
Since we consider a lattice model, the field operator here obeys the discrete bosonic commutation relations (\ref{eq:comm_psi}).
The second term in (\ref{eq:discrHam}) represents atomic interactions 
modeled by a purely on-site interaction
with a bare coupling constant on the lattice $g_0$. In practice,  
to recover the continuous space physics,
$l$ is taken to be smaller than both the healing length $\xi_{\rm heal}$ and the thermal de Broglie wavelength
$\lambda_{\rm dB}$. In the weakly interacting regime, $|a|\ll \xi_{\rm heal},\lambda_{\rm dB}$,
one can further take $l\gg |a|$ so that in the following we will identify $g_0$ with 
the effective coupling constant $g=4\pi \hbar^2 a/m$ where $a$ is the $s$-wave scattering length
\footnote{The exact relation between the bare coupling constant and the effective coupling constant
is $1/g=1/g_0+\int_{\mathrm{FBZ}}\frac{d^3k}{(2\pi)^3} \frac{m}{\hbar^2k^2}$, where FBZ=$[-\pi/l,\pi/l[^3$.}.
  
Initially at $t<0$, all the $N$ atoms are in the internal state $a$ in thermal equilibrium described by the 
canonical density operator
\be
\hat{\rho}=\frac{1}{Z}e^{-\beta \hat{H}_a}
\ee
with $\beta = 1/(k_BT)$ and $T \ll T_c$ where $T_c$ is the transition temperature for Bose-Einstein condensation.

At $t=0$ an electromagnetic pulse mixes the states $a$ and $b$. The pulse Hamiltonian acting during a time interval $t_{\rm pulse}$ is
\begin{equation}
\hat{V}_p=\frac{\hbar \Omega}{2i} \sum_{\bf r} dV \, ({\hat{\psi}}_{a}^\dagger {\hat{\psi}}_{b} - {\hat{\psi}}_{b}^\dagger {\hat{\psi}}_{a}) 
\end{equation}
In practice the timescale $t_{\rm pulse}$ is shorter than all the relevant timescales 
in the original Hamiltonians $H_{a,b}$ so that we can take the limit $t_{\rm pulse}\to 0$, $\Omega \to \infty$ with 
$\Omega t_{\rm pulse}=\pi/2$.
After integration of the Heisenberg equations of motion during $t_{\rm pulse}$, it is found that
the fields are transformed by the $\pi/2$ pulse as follows:
\begin{eqnarray}
{\hat{\psi}}_a(0^+)&=&\frac{1}{\sqrt{2}} [ {\hat{\psi}}_{a}(0^-) - {\hat{\psi}}_{b}(0^-) ] \label{eq:pulse1}\\
{\hat{\psi}}_b(0^+)&=&\frac{1}{\sqrt{2}} [ {\hat{\psi}}_{a}(0^-) + {\hat{\psi}}_{b}(0^-) ] \label{eq:pulse2} 
\end{eqnarray}
We are interested in the squeezing and quantum correlations that develop during the non-equilibrium dynamics following the pulse for $t>0$.

\subsection{Our Approach}
\label{sub:twofold}
The problem of the scaling of the squeezing for $N\to \infty$ in the multimode case implies the solution 
of the non-equilibrium quantum dynamics for a large number of atoms and a large number of modes. We cannot solve exactly this problem
even numerically. However, what can be solved exactly on a computer is the ``classical field equivalent" of our problem.
We then adopt the strategy summarized in Table~\ref{fig:table}. 
We use the classical field model to (i) perform numerical experiments and (ii) test a perturbative solution
that we can generalize to the quantum case. The quantum perturbative solution is then used
to get quantitative predictions on the real physical system.
\begin{table}[htb]
\caption{We cannot solve exactly the problem in the quantum case but we find an analytical perturbative solution. We check our perturbative approach in the classical case where we can solve the model exactly (numerically).}
\label{fig:table}
\begin{center}
\begin{tabular}{|c|c||c|c|}
\hline
Quantum field & solution & Classical field & solution \\
model         & available ? & model & available ? \\
\hline
& & & \\
$\hat{H} [{\hat{\psi}}_a({\bf r}),{\hat{\psi}}_b({\bf r})]$ & no & $H \left[ {{\psi}}_a({\bf r}),{{\psi}}_b({\bf r}) \right]$ & 
{yes} \\
& & & $\updownarrow$ \\
Perturbative  & {yes} & Perturbative  & {yes} \\
& & & \\
\hline
\end{tabular}
\end{center}
\end{table}

\subsection{Classical field model}
\label{sub:classical_field}
The classical field model \cite{kagan_svistunov_92,damle_kedar_96} is obtained by replacing the quantum fields with 
classical fields in the Hamiltonian \footnote{In this classical limit, the substitution $g_0\to g$ is required,
since the difference between $1/g_0$ and $1/g$ is due to quantum fluctuations.} 
\be
\hat{\psi}_a^\dagger, \hat{\psi}_a, \hat{\psi}_b^\dagger, \hat{\psi}_b \:\: \to \:\: {\psi}_a^\ast, {\psi}_a,  {\psi}_b^\ast, {\psi}_b.
\ee
In the equations of motion the commutators are then replaced by Poisson brackets.
The classical field model is useful when the interesting physics is given
by low-energy highly populated modes \cite{goral_gajda_01,davis_morgan_01,lobo_sinatra_04}.
For our classical field simulations, we assume that this is the case in the equilibrium state before the pulse, 
with all the particles in state $a$. The initial field ${\psi}_a^{(0)}$ then randomly samples the thermal equilibrium
classical field distribution for the canonical ensemble at temperature $T$
 \be
\rho_{\rm cl}=\frac{1}{Z}e^{-\beta H} \label{eq:rho_can_cl}
 \ee
 where $H$ is the classical Hamiltonian, which is a discrete version of the Gross-Pitaevskii energy functional:
 \begin{equation}
{H}=\sum_\kk \frac{\hbar^2 k^2}{2m} {a}^\ast_\kk {a}_\kk + 
		\frac{g}{2} dV \sum_{\rr}  
	\psi_a^\ast(\rr) \psi_a^\ast(\rr)
				\psi_a(\rr) \psi_a(\rr) 
\label{eq:E}
\end{equation}
 For the initially empty state $b$,  inspired by the Wigner quasi-probability distribution of the quantum density operator and
the truncated Wigner approach \cite{steel_olsen_98,sinatra_lobo_00,sinatra_lobo_01,sinatra_lobo_02} we represent the vacuum by a classical field ${\psi}_b^{(0)}$
 having  in each mode independent Gaussian complex fluctuations of zero mean and variance $1/2$:
More precisely, we set $b_{\bf k}^{(0)}=X+iY$ where the independent real random variables $X$ and $Y$ 
have the same Gaussian probability distribution
\be
P(x)=\sqrt{\frac{2}{\pi}} e^{- 2 x^2}
\ee
At $t=0$ the fields are mixed by the pulse according to (\ref{eq:pulse1})-(\ref{eq:pulse2}).
At later times, the fields, $\psi_a$ and $\psi_b$ evolve independently according to the 
discrete non-linear Schr\"odinger equation ($\nu=a,b$)
\begin{equation}
i \hbar \, \partial_t \psi_{\nu}= \left[ -\frac{\hbar^2 \Delta}{2m} 
+ g |\psi_{\nu}({\bf r},t)|^2  \right] \, \psi_{\nu} 
\label{eq:sc}
\end{equation}
where the discrete Laplacian $\Delta$ has the plane waves $\exp(i \kk\cdot\rr)$ on the lattice as eigenvectors
of eigenvalues $-k^2$. The lattice model automatically provides a momentum cut-off to the 
classical field model, corresponding to the boundaries of the first Brillouin zone FBZ=$[-\pi/l,\pi/l[^3$.
The various observables have a more or less pronounced dependence on the cut-off.
Here, guided by our analytical results (see section \ref{sec:analy}) we choose the cut-off such that, 
in the thermodynamic limit, the non-condensed density for an ideal gas in continuous space in the Bose condensed regime (zero chemical potential)
is exactly reproduced by the classical field model:
\be
\int_{\mathbb{R}^3} \frac{d^3k}{(2\pi)^3} \frac{1}{e^{\beta E_k}-1} = \int_{\mathrm{FBZ}} \frac{d^3k}{(2\pi)^3}  \frac{k_BT}{E_k}
\label{eq:cut_int}
\ee
where $E_k=\hbar^2 k^2/2m$ is the kinetic energy, the mode occupation numbers are given by the Bose formula
for the quantum case and by the equipartition formula for the classical case.
As revealed by the change
of integration variable ${\bf K}=\lambda_{\rm dB} {\kk}$,
the condition (\ref{eq:cut_int}) is an equation for $l/\lambda_{\rm dB}$ 
with $\lambda_{\rm dB}=[2\pi \hbar^2/(m k_BT)]^{1/2}$
the thermal de Broglie wavelength. 
The integrals on both sides of (\ref{eq:cut_int})
can be calculated analytically, see e.g.\ \cite{Pricoupenko:2007} for the integral in the right-hand side,
and the usual factor $\zeta(3/2)$ appears in the left-hand side (where $\zeta$ is the Riemann Zeta function).
The condition (\ref{eq:cut_int}) then gives
$E_k^{\rm max} \simeq 2.695 k_BT$ with $E_k^{\rm max}$ is the maximal kinetic energy on the grid,
here $E_k^{\rm max}=3\hbar^2 (\pi/l)^2/(2m)$
\footnote{In the classical field simulations, to ensure maximal ergodicity, we did not use a cubic lattice.
We used the following aspect ratios for the quantization box,  $L_x^2:L_y^2:L_z^2=\sqrt{2}:(1+\sqrt{5})/2:\sqrt{3}$,
and we discretized the positions with the same even number of points $n_{\rm max}$ along each direction.
This corresponds in condition (\ref{eq:cut_int}) 
to a slightly asymmetric FBZ.}.

\section{Numerical Experiments}
\label{sec:numexp}

\subsection{Dimensional analysis}
\label{subsub:dim}
As specified in subsection \ref{sub:model}, when the lattice model approaches the continuous space physics for our observable (the spin squeezing), the physical parameters of the model 
are the atom mass $m$, the effective coupling constant $g$ characterizing low energy binary interactions between the atoms, 
the temperature $T$, the total atom number $N$ and system volume $V$:
\be
{\hbar^2 \over m}\,, \:\:\: g\,, \:\:\: k_BT\,, \:\:\: N \,, \:\:\: V  \label{eq:params}
\ee
The spin squeezing parameter optimized over time, $\xi_{\rm min}^2$, is a dimensionless quantity. 
It is therefore a function of the independent dimensionless combinations that we can form from the ensemble (\ref{eq:params}):
\be
\xi^2_{\rm min} = f (N, \sqrt{\rho a^3}, {k_B T\over \rho g}) 
\label{eq:form_gen}
\ee
Here $\sqrt{\rho a^3}$ is the ``small parameter" such that $\sqrt{\rho a^3}\ll 1$ characterizes the weakly interacting limit, and
$\rho g$ is the mean field chemical potential of the gas (at $T=0$). The same dimensional analysis and
the same general form of $\xi^2_{\rm min}$ hold for the classical field model.

\subsection{Existence of a thermodynamic limit for $\xi^2$}
\label{subsub:TL}
\begin{figure}[t!]
\begin{center}
\resizebox{0.75\columnwidth}{!}{\includegraphics{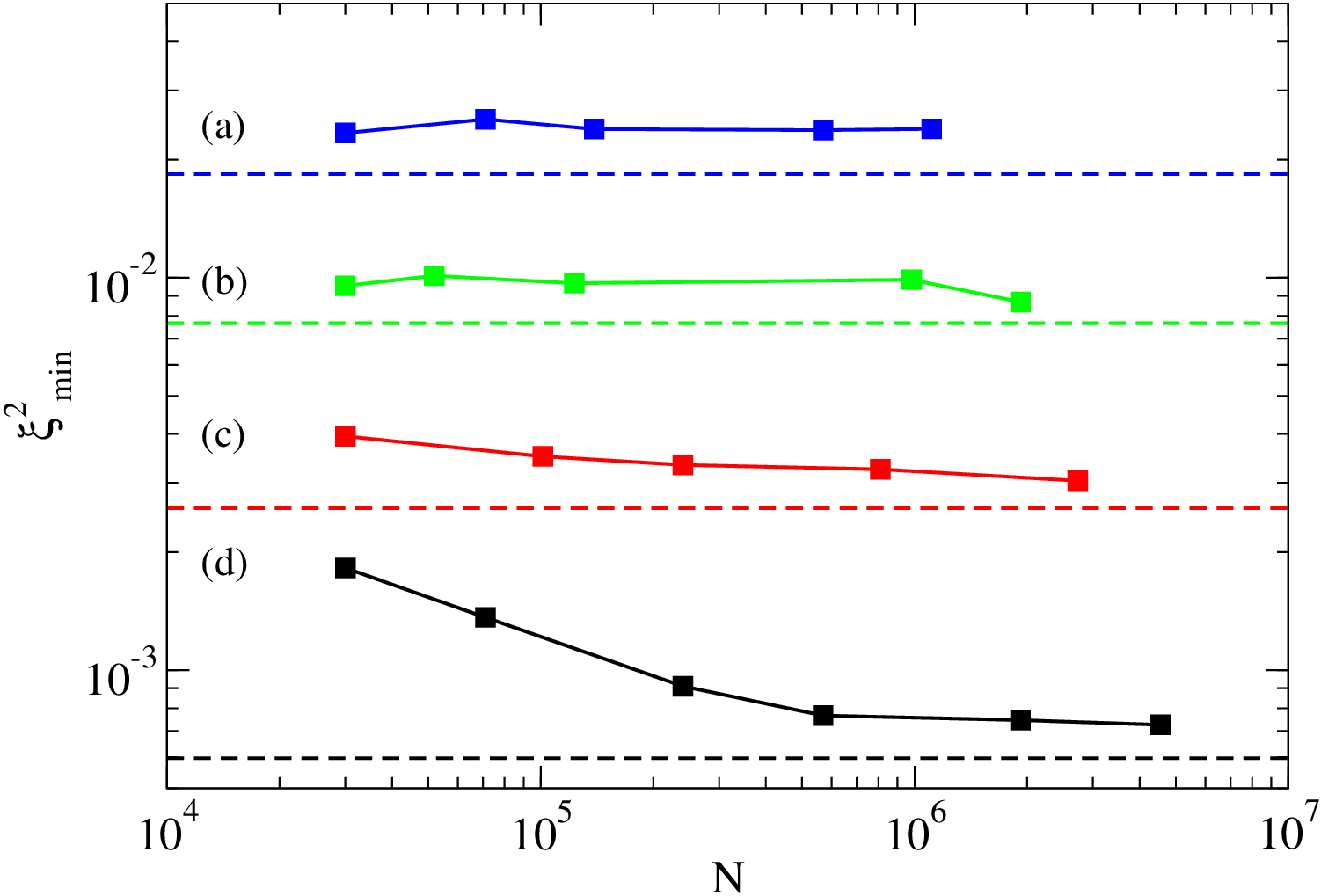}}
\caption{(Color online).\ Minimal squeezing parameter (that is, minimized over time) for four different temperatures and increasing system sizes
in the thermodynamic limit $N\to \infty$, $\rho,g,T=\mbox{\rm constant}$. 
Squares: classical field simulations result. 
$k_BT/\rho g=1.13(a), 0.78(b), 0.50(c), 0.28(d)$.
For all the points $\sqrt{\rho a^3}=1.32\times10^{-2}$. The horizontal dashed lines are analytical results 
in the thermodynamic and weakly interacting limit (\ref{eq:xibest}) that are the classical field equivalent of (\ref{eq:xibest_q}).
The grid sizes (number of points per direction) are $n_{\rm max}=12,16,20,32,36$ for (a), $n_{\rm max}=10,12,16,32,40$ for (b),
$n_{\rm max}=12,16,24,36,40$ for (c) and $n_{\rm max}=6,8,12,16,24,32$ for (d).
\label{fig:TL}}
\end{center}
\end{figure}

We have performed classical field simulations,
increasing the system size in the thermodynamic limit
\be
N \to \infty\,; \hspace{0.2cm} V\to \infty\,;  \hspace{0.5cm} \rho,\; g,\; T\; =\; {\rm constant}\,,
\label{eq:def_lim_therm}
\ee
In Fig.\ref{fig:TL} we show the result for four different temperatures. 
The squeezing parameter, minimized over time, converges to a finite value. According to the general form
(\ref{eq:form_gen}), $\xi_{\rm min}^2$ then depends on $k_BT/\rho g$ and $\sqrt{\rho a^3}$.
The first parameter $k_BT/\rho g$ is varied in Fig.\ref{fig:TL}, whereas
the parameter $\sqrt{\rho a^3}$ defining the weakly interacting regime is maintained constant in that figure. 
Note that for curves (a)-(c) the limit is already almost reached for $N=3\times10^4$
while a larger system is needed for the lowest temperature curve (d).

\subsection{Weakly interacting limit of $\xi^2$}
\label{subsub:CD}
\begin{figure}[t!]
\begin{center}
\resizebox{0.75\columnwidth}{!}{\includegraphics{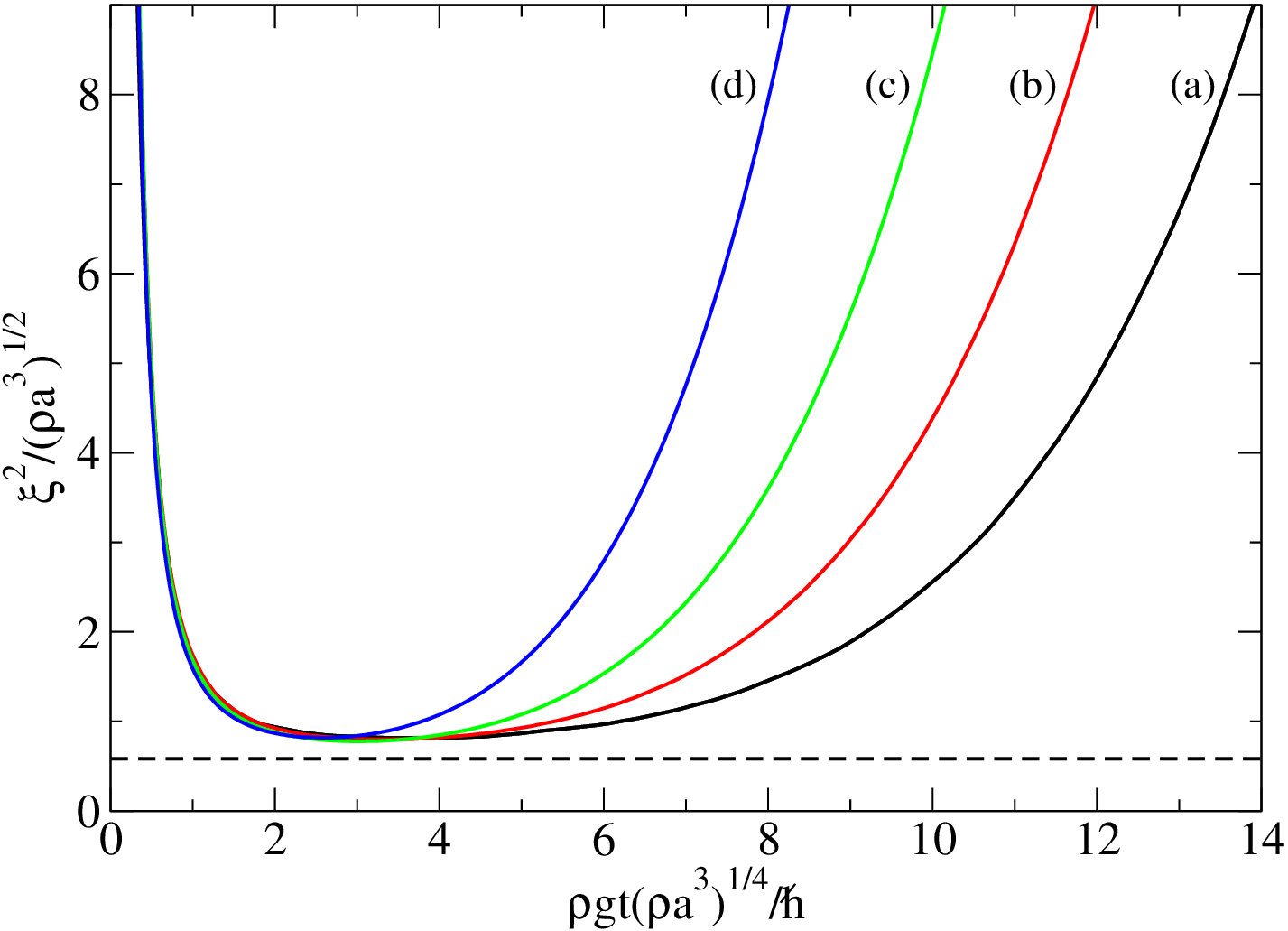} }
\caption{(Color online).\ Squeezing parameter $\xi^2$ divided by $\sqrt{\rho a^3}$ as a function of a rescaled time, in the weakly interacting limit: 
$\rho \to \infty$, $g \to 0$ with $\rho g,T={\rm constant}$. Solid lines: classical field simulations. Parameters: $k_BT/\rho g=0.5$
and $n_{\rm max}=32$  for all the curves. 
(a) $N=3\times10^4$, $\sqrt{\rho a^3}=1.32\times10^{-2}$; 
(b) $N=10^5$, $\sqrt{\rho a^3}=3.96\times10^{-3}$;
(c) $N=3\times10^5$, $\sqrt{\rho a^3}=1.32\times10^{-3}$; 
(d) $N=6\times10^5$, $\sqrt{\rho a^3}=6.59\times10^{-4}$. Horizontal line: analytical result (\ref{eq:xibest}).
\label{fig:CD}}
\end{center}
\end{figure}

Starting from a point that is already at the thermodynamic limit for each temperature, 
we have performed other simulations, going more deeply
in the weakly interacting limit
$\rho \to \infty$, $g\to 0$, $\rho g,T={\rm constant}$ \cite{Castin:1998} 
\footnote{In our simulations (with finite size systems) 
we increase $N$ and decrease $g$ while $V$, $T$ and $Ng$ are fixed.}. 
In this limit the small parameter  $\sqrt{\rho a^3}$ tends to zero.
In Fig.\ref{fig:CD}, for a fixed value of $k_BT/\rho g$, we show the squeezing parameter
divided by $\sqrt{\rho a^3}$ as a function of a rescaled time, for various values of $\sqrt{\rho a^3}$.
It is apparent  that the rescaled minimal squeezing is nearly constant in the figure.
For weak interactions, $\xi_{\rm min}^2/\sqrt{\rho a^3}$ is thus a function of $k_B T/\rho g$ only.

\subsection{Spin squeezing as a function of the temperature}
\begin{figure}[tb]
\begin{center}
\resizebox{0.75\columnwidth}{!}{\includegraphics{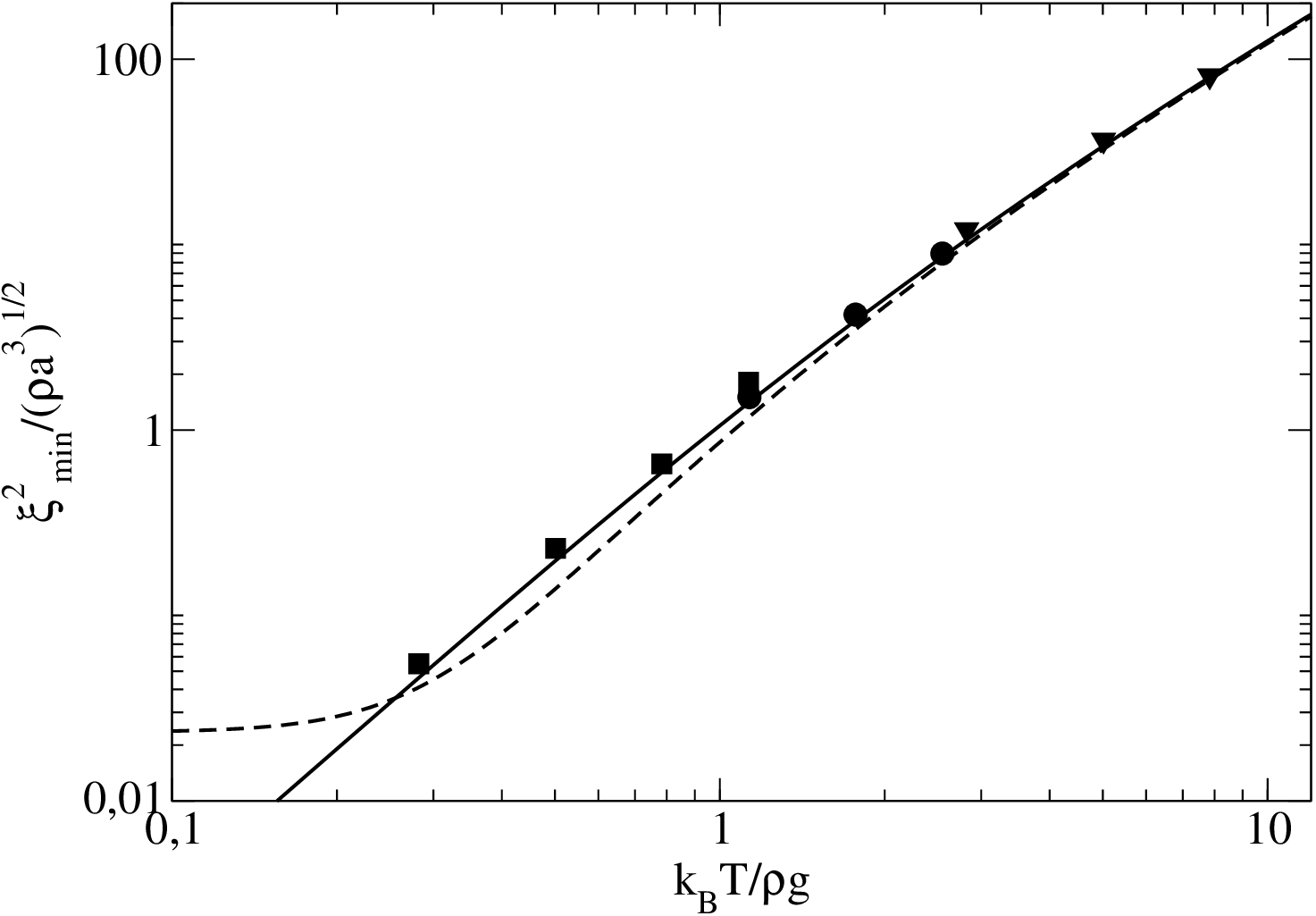} }
\caption{Minimal squeezing parameter $\xi_{\rm min}^2$ divided by $\sqrt{\rho a^3}$ as a function of $k_BT/\rho g$.
Symbols: classical field simulations with $\sqrt{\rho a^3}=1.32 \times10^{-2} $ (squares), $1.94 \times10^{-3}$
(disks), and $\sqrt{\rho a^3}=4.17 \times10^{-4} $ (triangles). Solid line: 
Analytical classical field result (\ref{eq:xibest}). Dashed line: Quantum result (\ref{eq:xibest_q}). 
The grid sizes, for increasing $k_BT/\rho g$ are $n_{\rm max}=32,36,40,36$ for squares, $n_{\rm max}=24,32,32$ for disks and
$n_{\rm max}=32$ for triangles. The initial non condensed fractions in component $a$ before the pulse are 
$\langle N_{\rm nc} \rangle/N=0.01,0.02,0.04,0.07$ for squares, $\langle N_{\rm nc} \rangle/N=0.01,0.02,0.04$ for disks and
$\langle N_{\rm nc} \rangle/N=0.01,0.02,0.05$ for triangles.
\label{fig:xibest}}
\end{center}
\end{figure}

From the numerical experiments described in \ref{subsub:TL} and \ref{subsub:CD}, we conclude that, in the double thermodynamic plus weakly interacting limit, the squeezing parameter minimized over time is equal to $\sqrt{\rho a^3}$ times a universal function of $k_BT/\rho g$:
\be
\xi^2_{\rm min}=\sqrt{\rho a^3} \; F\left({k_BT \over \rho g}\right)  \label{eq:univF}
\ee
We show the universal scaling of the minimal squeezing parameter in the thermodynamic limit in Fig.\ref{fig:xibest}, where
we collect the results of several simulations for different temperatures and interaction strengths.  
In the following section we shall develop the analytical theory that gives the explicit expression of the function $F$ appearing in (\ref{eq:univF}). 
The analytical result  for the classical field theory is represented as a full line in Fig.\ref{fig:xibest}. It reads:
\begin{equation}
\xi_{\rm min}^2 =   
\int_{\rm FBZ}\! {d^3 k \over (2\pi)^3} \, \frac{s_k^4}{2\rho} \, n_k^{(0)} \left(  {(s_k^{(0)} )^2 \over  s_k^4 }  + {s_k^4 \over (s_k^{(0)})^2 }  \right)
\label{eq:xibest}
\end{equation}
Here $\rho=N/V$ is the total density, $s_k=U_k+V_k$, $s_k^{(0)}=U_k^{(0)}+V_k^{(0)}$ are Bogoliubov functions defined in (\ref{eq:sk}) and (\ref{eq:sk0}), $n_k^{(0)}=k_BT/\epsilon_k^{(0)}$ are the equilibrium occupation numbers of Bogoliubov modes before the pulse 
with $\epsilon_k^{(0)}=[E_k(E_k+2\rho g)]^{1/2}$ and the integral is restricted to the first Brillouin zone $\mathrm{FBZ}$ 
with $k_\nu \in [-k_{\nu}^{\rm max},k_{\nu}^{\rm max}[$ for the three directions $\nu=x,y,z$. 
We note a very good agreement between the analytics and the simulations for all the points.
For comparison, we also show the analytical result for the quantum field (\ref{eq:xibest_q}), in the zero-lattice spacing limit,
as a dashed line.
Note how well the classical results with the cut-off prescription (\ref{eq:cut_int}) reproduce the quantum analytical results for
$k_BT > \rho g$.

\section{Analytical results}
\label{sec:analy}
We want here to address the fully quantum case and calculate analytically the function $F(k_BT/\rho g)=\xi^2_{\rm min}/\sqrt{\rho a^3}$ 
of (\ref{eq:univF}) using Bogoliubov theory. In \ref{sub:bogol} we present a modulus-phase 
reformulation of the Bogoliubov theory generalized for a bimodal condensate; in \ref{sub:expansion} and \ref{subsec:rotemfx2}
we derive the expansion of the spin squeezing parameter in the thermodynamic limit, and we discuss the final analytical results in \ref{subsec:results}.

\subsection{Modulus-phase Bogoliubov formalism for bimodal condensates}
\label{sub:bogol}

We generalize here to two-components the $U(1)$-symmetry preserving Bogoliubov theory of \cite{Castin:1998},
see also \cite{Sorensen:2001}. 
As spin squeezing in bimodal condensates is due to phase dynamics, we rephrase
the theory in terms of the relative phase operator of the $a$ and $b$ condensates. This is a crucial step that allows us to obtain
results by a perturbative expansion in powers of that relative phase.
The relative phase is simply the difference of the condensate phases $\hat{\theta}_{a,b}$
introduced as hermitian operators conjugate to the condensate atom numbers
\cite{Girardeau:1959,Nieto:1968}. This is a valid description in the subspace excluding the vacuum state (no particles) 
for the condensate modes, which is sufficient here: 
as we suppose initially $T\ll T_c$, both condensates are highly populated 
after the mixing pulse with $\langle N_a \rangle=\langle N_b \rangle=N/2$.
Some useful expressions involving the Bogoliubov amplitudes and some commutation relations are given in Appendix \ref{app:useful}.
 
Within each atomic internal state $a$ or $b$, we split the bosonic field operator into the condensate and non condensed modes contributions:
\be
\hat{\psi}_a=\frac{\hat{a}_{\bf 0}}{\sqrt{V}}+\hat{\psi}_{a \perp}\,, \hspace{0.3cm} \hat{\psi}_b=\frac{\hat{b}_{\bf 0}}{\sqrt{V}}+\hat{\psi}_{b \perp} \label{eq:split}
\ee
For the annihilation operators in the condensate modes we introduce the modulus-phase representation
\begin{eqnarray}
\label{eq:mpra}
\hat{a}_{\bf 0}&=&e^{i {\hat{\theta}}_a} \sqrt{{\hat{N}}_{a{\bf 0}} } \:\: \:\: ,  \:\:\:\: [{\hat{N}}_{a{\bf 0}},{\hat{\theta}}_{a}]=i \\
\hat{b}_{\bf 0}&=&e^{i {\hat{\theta}}_b} \sqrt{{\hat{N}}_{b{\bf 0}} } \:\: \:\: ,  \:\:\:\: [{\hat{N}}_{b{\bf 0}},{\hat{\theta}}_{b}]=i \,,
\label{eq:mprb}
\end{eqnarray}
while for the non condensed modes we introduce the 
number conserving fields \cite{Castin:1998,Gardiner:1997}
\be
\hat{\Lambda}_a=e^{-i\hat{\theta}_a}\hat{\psi}_{a \perp} \hspace{0.15cm} ; \hspace{0.15cm}  \hat{\Lambda}_b\!=e^{-i\hat{\theta}_b}\hat{\psi}_{b \perp} 
\label{eq:Lambdas}
\ee
that we expand over Bogoliubov modes with amplitudes $\hat{c}_{a\kk}$ and $\hat{c}_{b\kk}$.
To perform this expansion, one has to distinguish the time before the pulse and the time after the pulse.

{\it Before the pulse}, at time $t=0^-$, all the $N$ atoms are in the internal state $a$, with a zero-temperature mean field chemical 
potential $\mu=\rho g$, and we expand the number conserving field $\hat{\Lambda}_a(t=0^-)$ over the Bogoliubov modes:
\be
\hat{\Lambda}_a^{(0)}= \!\! \sum_{\kk \neq {\rm 0}}  
\left[ U_k^{(0)} \hat{c}_{a\kk}^{(0)}+V_k^{(0)} 
(\hat{c}_{a -\kk}^{(0)})^\dagger \right] \frac{e^{i\kk \cdot \rr}}{\sqrt{V}} \label{eq:Lambda0} 
\ee
Here the exponent ${}^{(0)}$ over the operators indicates that the operators are considered at time $t=0^-$. 
The bosonic operator $\hat{c}_{a\kk}^{(0)}$ annihilates a Bogoliubov quasi-particle of wave vector $\kk$, and
eigenenergy
\be
\epsilon_k^{(0)}=[E_k(E_k+2\rho g)]^{1/2}
\ee
where we recall that $E_k=\hbar^2 k^2/(2m)$ is the kinetic energy contribution. 
The corresponding amplitudes of the Bogoliubov mode, correctly normalized as $[U_k^{(0)}]^2-[V_k^{(0)}]^2=1$, are given by
\be
s_k^{(0)} \equiv U_k^{(0)}+V_k^{(0)}=\frac{1}{U_k^{(0)}-V_k^{(0)}}=\left( {E_k \over E_k + 2\rho g}\right)^{1/4}  \label{eq:sk0}
\ee
Before the pulse, the field in the $b$ state is in the vacuum state, so the modal expansion is performed
over the usual single particle plane wave eigenbasis.

{\it After the pulse}, at $t\ge 0^+$, the particles are on average equally distributed among the two internal
states $a$ and $b$, $\langle {N}_a \rangle = \langle {N}_b \rangle = N/2$. There are now two condensates, with interaction constants 
$g_{aa}=g_{bb}=g$ among atoms of same internal state, and no interaction ($g_{ab}=0$) among atoms of different internal states.
Similarly to (\ref{eq:Lambda0}), we perform after the pulse the modal decomposition of the number conserving fields
in each internal state $\sigma=a,b$:
\be
\hat{\Lambda}_\sigma = \sum_{\kk \neq {\rm 0}}  
\left[ U_k \hat{c}_{\sigma\kk}+V_k \hat{c}_{\sigma -\kk}^\dagger \right] \frac{e^{i\kk \cdot \rr}}{\sqrt{V}} \label{eq:Lambdat} \\
\ee
where $\hat{c}_{\sigma \kk}$ annihilates a Bogoliubov quasi-particle of wave vector $\kk$ in internal state $\sigma$.
The Bogoliubov mode amplitudes $U_k$, $V_k$ and the eigenenergies $\epsilon_k$ do not depend on the internal state
and are deduced from the ones at $t=0^-$ by replacing the mean field term $\rho g$ by $\rho g/2$:
\bea
s_k &\equiv& U_k+V_k=\frac{1}{U_k-V_k}=\left( {E_k \over E_k + \rho g}\right)^{1/4} \label{eq:sk} \\
\epsilon_k &=& [E_k(E_k+\rho g)]^{1/2} \,, \label{eq:epsk}
\eea
Note that this involves an approximation: In principle, the Bogoliubov modes in internal state $\sigma=a$ or $b$ depend
on the actual number of particles $N_\sigma$ in that state, which has small $\approx 1/\sqrt{N}$ relative fluctuations since,
after the pulse, ${N}_\sigma$ has a binomial distribution peaked around $N/2$.
Taking into account this effect changes the mathematical structure of the theory, since
the Bogoliubov amplitudes $U_k$ and $V_k$, and thus the quasi-particle annihilation operators
$\hat{c}_{\sigma\kk}$, would then depend on the total number operator $\hat{N}_\sigma$ in state $\sigma$,
which is beyond the scope of the present work. 
We nevertheless verified numerically on the complete Bogoliubov theory (in the classical field model)
that this fixed-Bogoliubov-mode approximation is extremely accurate both at short and long times,
introducing (for the typical parameters considered in our figures) a relative error on $\xi^2$  lower than $10^{-2}$ comparable to
our statistical error bars with $10^5$ realizations.
Another important point is that the Bogoliubov quasi-particles are not at thermal equilibrium after the pulse,
so that, for example, their mean occupation numbers are not given by the Bose formula.
At the level of the Bogoliubov approximation, the quasi-particles do not interact and cannot thermalize, the corresponding
quasi-particle creation operators evolve in Heisenberg picture with simple phase factors:
\be
\hat{c}_{\sigma \kk}(t) \underset{t>0}{=} e^{-i\epsilon_k t/\hbar} \hat{c}_{\sigma \kk}(0^+)
\label{eq:ckt}
\ee
for $\sigma=a,b$.  The validity of this no-thermalization approximation is discussed in subsection \ref{subsec:results}.

To express the evolution of the particle annihilation operators $\hat{a}_{\oo}$ and $\hat{b}_\oo$ in the condensate
modes, we use the modulus-phase representation (\ref{eq:mpra}, \ref{eq:mprb}). For the modulus, one simply
uses the conservation of the total atom number in each internal state, 
\be
\hat{N}_{\sigma\oo} = \hat{N}_\sigma -\hat{N}_{\sigma\perp} = \hat{N}_\sigma - \sum_\rr dV \hat{\Lambda}_\sigma^\dagger \hat{\Lambda}_\sigma
\label{eq:conservation}
\ee
so that $\hat{N}_{\sigma\oo}$ can be expressed in terms of the quasi-particle operators $\hat{c}_{\sigma\kk}$.
For the phase, we use within each internal state $\sigma=a,b$ the equation of motion derived for a single component
in \cite{Sinatra:2007} and truncated at the level of the Bogoliubov approximation:
\be
\frac{d}{dt} \hat{\theta}_\sigma  =  - \chi \left( {\hat{N}}_{\sigma}-\frac{1}{2} \right) -\frac{\chi}{2}
\sum_{\rr} dV \left( {\hat{\Lambda}_\sigma}^2 + 2{\hat{\Lambda}_\sigma}^\dagger {\hat{\Lambda}_\sigma} + 
{\hat{\Lambda}_\sigma}^{\dagger 2} \right) 
\label{eq:thetadotsigma}
\ee
where we have introduced $\chi\equiv g/(\hbar V)$. Replacing the operators $\hat{\Lambda}_\sigma$ by their modal
expansion (\ref{eq:Lambdat}), one gets contributions that do not oscillate in time, and contributions such
as $\hat{c}_{\sigma\kk} \hat{c}_{\sigma-\kk}$ that oscillate in time (at the frequency $2\epsilon_k/\hbar$ in the example).
As we shall need the value of the phase operators at long times $t$ (typically $\epsilon_k t/\hbar \gg 1$)
rather than the value of its derivative,
we argue as in \cite{Sinatra:2007} that the oscillating terms in (\ref{eq:thetadotsigma}), after temporal integration,
give a negligible contribution to the squeezing parameter. 
We have checked this approximation analytically: The expression for $\xi^2$ fully including
the oscillating terms in the phase difference operator is given in the Appendix \ref{app:lourd}, it
corrects the approximate expression (\ref{eq:central}) of $\xi^2$ by typically a sub-percent effect at intermediate times, 
and by a vanishing amount at large times.  These oscillating terms in $\hat{\theta}_a-\hat{\theta}_b$
are thus of little physical relevance for the spin squeezing.
Keeping only the non-oscillating terms gives for the relative phase operator of the two condensates at time $t$:
\bea
\!\!\! (\hat{\theta}_a-\hat{\theta}_b)(t)&=&(\hat{\theta}_a-\hat{\theta}_b)(0^+) 
-\chi t \left[ \hat{N}_a-\hat{N}_b  + \hat{\cal D} \right] \label{eq:rephase} \\
\!\!\! \hat{\cal D} &=& \sum_{{\bf k}\neq{\bf 0}} (U_k+V_k)^2 (\hat{n}_{a{\bf k}}-\hat{n}_{b{\bf k}})(0^+)  \label{eq:calD}
\eea
where we have introduced the quasi-particle number operators $\hat{n}_{\sigma \kk}=\hat{c}_{\sigma\kk}^\dagger \hat{c}_{\sigma\kk}$,
which are constants of motion in the Bogoliubov approximation, see (\ref{eq:ckt}).
The multimode contribution $\hat{\cal D}$ (\ref{eq:calD}) to the relative phase (\ref{eq:rephase}) will play a central role
in what follows. It is indeed because of this term (neglected in the usual two-mode models)
that the squeezing parameter is bounded from below by a non-zero value in the thermodynamic limit.

The last step is to relate the various operators at time $t=0^+$ to their values just before the pulse.
Since the state of the system is known at $t=0^-$, this fully specifies the ``initial" conditions for
the time evolution of the operators after the pulse.
The derivation and the more precise results are given in the Appendix \ref{app:initial_phase}.
Here we give the main conclusions.  The initial value of the condensate phase difference is, in the
large $N$ limit:
\be
\label{eq:init_diff_theta}
(\hat{\theta}_a-\hat{\theta}_b)(0^+) = \frac{i}{2} \left\{
\left(\hat{N}_{a{\bf 0}}^{(0)}\right)^{-1/2} 
\!\!\!\!, 
\left(\tilde{b}_{\bf 0}^{(0)}-\mbox{h.c.}\right)
\right\}
+O\left(\!\frac{1}{N^{3/2}}\!\right)
\ee
where $\tilde{b}_\oo^{(0)}\equiv e^{-i\theta_a^{(0)}} \hat{b}_{\bf 0}^{(0)}$ and $\{\ ,\ \}$ stands for the anticommutator.
This results from the coherent mixing of the initial condensate amplitude with the vacuum noise fluctuations 
in the initially empty internal state $b$, in the same spatial mode as the initial condensate in state $a$, 
that is in the plane wave with zero wave vector.
Similarly, the quasi-particle annihilation operators just after the pulse are coherent superpositions of the initial field
fluctuations, mainly thermal fluctuations $\hat{A}$ in $a$ and only vacuum fluctuations $\hat{B}$ in $b$.
In the large $N$ limit, 
\be
c_{\sigma \kk} (0^+) = \frac{\hat{A}_\kk \mp \hat{B}_\kk}{\sqrt{2}} + O\left(\frac{1}{N^{1/2}}\right)
\label{eq:initczp}
\ee
where the $-$ sign is for $\sigma=a$ and the $+$ sign is for $\sigma=b$. The expression of $\hat{A}_\kk$ in terms
of $t=0^-$ operators of the $a$ internal state, and the expression of $\hat{B}_\kk$ in terms of
$t=0^-$ operators of the $b$ internal state, naturally appear in the calculations of Appendix \ref{app:initial_phase}:
\bea
\label{eq:defA}
\hat{A}_\kk & \equiv & (U_k U_k^{(0)}- V_k V_k^{(0)})\, c_{a\kk}^{(0)} + (U_k V_k^{(0)} - V_k U_k^{(0)})\, c_{a-\kk}^{(0)\dagger} 
\\
\hat{B}_{\kk} &\equiv & U_k e^{-i\hat{\theta}_a^{(0)}} \hat{b}_{\kk}^{(0)} 
- V_k e^{i\hat{\theta}_a^{(0)}} \hat{b}_{-\kk}^{(0)\dagger} 
\label{eq:defB}
\eea
In Appendix \ref{app:GammaDelta}, it is pointed out that these number conserving operators obey bosonic commutation relations
and all their second moments are explicitly evaluated.

\subsection{Double expansion method in the thermodynamic and weakly interacting limit}
\label{sub:expansion}

We shall now apply the Bogoliubov theory developed in section \ref{sub:bogol}  to the calculation of the time-dependent
squeezing parameter $\xi^2$ defined in (\ref{eq:xi2def}).
For the configuration that we consider, symmetric under the exchange of $a$ and $b$, 
the mean spin is always aligned along $x$. The minimum transverse spin variance is then
\begin{equation}
\label{eq:dspm}
\Delta S_{\perp,{\rm min}}^2={1\over 2}\left[ \langle \hat{S}_y^2 \rangle+\langle \hat{S}_z^2 \rangle 
-\sqrt{(\langle \hat{S}_y^2 \rangle-\langle \hat{S}_z^2\rangle )^2+\langle \{\hat{S}_z,\hat{S}_y\}\rangle^2} \right] \,,
\end{equation}
where $\{\ ,\ \}$ is the anticommutator. From the definition (\ref{eq:Sz})
it appears that $\hat{S}_z$ is a constant of motion,
its variance can thus be evaluated just after the pulse, at $t=0^+$. According to (\ref{eq:pulse1},\ref{eq:pulse2}),
the pulse applies to the collective spin a rotation of angle $\pi/2$ around $y$ axis, so that
\be
\hat{S}_z(t>0)=-\hat{S}_x^{(0)} \hspace{5mm}\mbox{and}\hspace{5mm}  \langle \hat{S}_z^2\rangle = \frac{N}{4}
\label{eq:varsz}
\ee
To obtain the spin variance $\langle \hat{S}_y^2 \rangle$ and the spin correlation $\langle \{\hat{S}_y,\hat{S}_z\} \rangle$,
the challenge is to determine, as a function of time, the operator $\hat{S}_y$, or equivalently the antihermitian
part of the operator
$\hat{S}_+$ introduced in (\ref{eq:SxSy})  (in its discrete version for the lattice model),
since $\hat{S}_y=(\hat{S}_+-\mbox{h.c.})/(2i)$.
In the expression of $\hat{S}_+$, one applies the splitting (\ref{eq:split}) of the bosonic fields in
the condensate and the non-condensed contributions, one uses the modulus-phase representation 
(\ref{eq:mpra}, \ref{eq:mprb}) for the condensate part and one introduces the number-conserving fields
$\hat{\Lambda}_\sigma$ for the non-condensed  part, to obtain:
\bea
\hat{S}_+&=&e^{-i(\hat{\theta}_a-\hat{\theta}_b)} \left(\frac{N}{2}+\hat{F}\right) 
\hspace{0.5cm} {\rm with} \hspace{0.5cm} \label{eq:defSplus}\\
\hat{F}&=&\sqrt{(\hat{N}_{a {\bf 0}}+1) \hat{N}_{b {\bf 0}} }-\frac{N}{2} 
+ \sum_{\bf r} dV \hat{\Lambda}_a^\dagger \hat{\Lambda}_b
\label{eq:F}
\eea

Guided by the numerical experiments in section \ref{sec:numexp}, 
we have developed a systematic {\it double expansion} technique to determine 
$\langle \hat{S}_y^2 \rangle$ and $\langle \{\hat{S}_y ,\hat{S}_z\} \rangle$. 
The two small parameters controlling the large system size limit [i.e. the thermodynamic limit (\ref{eq:def_lim_therm})] 
and the Bogoliubov limit are
\be
\epsilon_{\rm size} = \frac{1}{N} \hspace{5mm} \mbox{and}  \hspace{5mm} \epsilon_{\rm Bog} \equiv {\langle N_{\rm nc} \rangle \over N}
\ee
where $\langle N_{\rm nc}\rangle = \langle \hat{N}_{a\perp}^{(0)}\rangle$ 
is the mean number of non-condensed particles in the initial state, which is
indeed much smaller than $N$ for a weakly interacting gas at $T\ll T_c$.
For the Bogoliubov expansion, we will keep terms up to order one included in $\epsilon_{\rm Bog}$; keeping higher order terms
would not be consistent with the use of the quadratic Bogoliubov Hamiltonian.
To determine the required order of the large system size expansion, we note that $\langle \hat{S}_x\rangle$ in the denominator of
(\ref{eq:xi2def}) remains close to its $t=0^+$ value $N/2$ over the relevant time scales 
(that are finite in the thermodynamic limit with $\rho g t/\hbar \ll N^{1/2}$), so that
$\xi^2 \approx 4 \Delta S_{\perp,\mathrm{min}}^2/N$.
To have a vanishingly small error on $\xi^2$ in the thermodynamic limit,
we will keep in $\langle \hat{S}_y^2 \rangle/N$ and $\langle \{\hat{S}_y, \hat{S}_z\} \rangle/N$ terms up to order
zero included in $\epsilon_{\rm size}$, that is we can neglect the contributions that
tend to zero when $\epsilon_{\rm size}\to 0$.

The systematic technique to determine the order of an operator is to estimate its mean value
and its standard deviation in the quantum state of the system. This is safer than a simple guess,
in particular when the operator has a vanishing expectation value. A relevant example is the
operator $\hat{\cal D}$ defined in (\ref{eq:calD}), that will play a crucial role
in the best achievable squeezing. After a superficial look at (\ref{eq:calD}), one may believe
that $\hat{\cal D}$ scales as $N$ in the thermodynamic limit, since it involves a sum over all
modes, and that it scales as $\epsilon_{\rm Bog}$ in the Bogoliubov limit since it involves
the quasi-particle number operators $\hat{n}_{\sigma\kk}$. However, it is actually the differences
$\hat{n}_{a\kk}-\hat{n}_{b\kk}$ that matter, and for the particular state resulting from
a $\pi/2$ pulse applied on a gas initially in the $a$ internal state.
As a consequence, the expectation value of $\hat{\cal D}$ is zero, it is the variance of $\hat{\cal D}$
which scales as $N$, so $\hat{\cal D}$ scales as $N^{1/2}$ in the thermodynamic limit.
To determine its scaling with $\epsilon_{\rm Bog}$, we
keep the leading term (\ref{eq:initczp}) in the value of the quasi-particle annihilation
operator just after the pulse, to obtain
\be
\hat{\cal D}\simeq - \sum_{\kk\neq \oo} (U_k+V_k)^2 (\hat{A}_\kk^\dagger \hat{B}_\kk+ \hat{B}_\kk^\dagger \hat{A}_\kk)
\label{eq:Dut}
\ee
This scales with $N$ as $N^{1/2}$ since each term has a zero mean.
Intuitively, this scales with $\epsilon_{\rm Bog}$ as $\epsilon_{\rm Bog}^{1/2}$: $\hat{B}_\kk$ is of order unity since it corresponds
to vacuum field fluctuations in the initial empty state $b$, and $\hat{A}_\kk$ is of order
$\epsilon_{\rm Bog}^{1/2}$ since it corresponds to the initial non-condensed field fluctuations in state $a$.
We thus conclude that
\be
\hat{\cal D} \approx (N \epsilon_{\rm Bog})^{1/2}
\label{eq:odg_D}
\ee
This is confirmed by the correlation functions of $\hat{A}$ and $\hat{B}$ given in the
Appendix \ref{app:GammaDelta}, that allow an explicit calculation of  $\langle \hat{\cal D}^2\rangle/N$,
which is indeed $\approx \epsilon_{\rm Bog}$, see Eqs.(\ref{eq:xi2min_d},\ref{eq:xibest_q}).

The same analysis is applied to the various operators appearing in the antihermitian part of (\ref{eq:defSplus}),
writing for simplicity $\hat{F}=\hat{F}_R + i \hat{F}_I$, where $\hat{F}_R$ and $\hat{F}_I$ are hermitian.
It is found in the Appendix \ref{app:details} that
\bea
\hat{\theta}_a-\hat{\theta}_b &\approx& \frac{1}{N^{1/2}} \\
\hat{F}_I &\approx& (N \epsilon_{\rm Bog})^{1/2} \\
\hat{F}_R &\approx& N \epsilon_{\rm Bog} \pm (N \epsilon_{\rm Bog})^{2/3}
\label{eq:F_R}
\eea
Contrarily to the first two operators, $\hat{F}_R$ has a non-zero expectation value,
and the writing of its estimate in (\ref{eq:F_R}) corresponds to its mean value plus or minus its standard deviation. It turns out that the fluctuations of
$\hat{F}_R$ are negligible so that the operator can be replaced
by its mean value. The weak value of the phase difference operator shows that
its exponential in (\ref{eq:defSplus}) can be expanded to first order, 
which substantially simplifies the calculations.
The final result, up to zeroth order included in $\epsilon_{\rm size}$
and up to first order included in $\epsilon_{\rm Bog}$, is
\bea
\langle \hat{S}_x \rangle &=& \frac{N}{2} \left( 1 + 2 \frac{\langle \hat{F}_R\rangle}{N} \right) 
\label{eq:finalsx} \\
\frac{\langle \hat{S}_y^2\rangle}{N} &=& 
\frac{\langle \hat{F}_I^2\rangle}{N} 
+\langle (\hat{\theta}_a-\hat{\theta}_b)^2\rangle \left(\frac{N}{4}
+\langle \hat{F}_R\rangle\right) 
 -\frac{1}{2} \langle \{ \hat{F}_I, \hat{\theta}_a-\hat{\theta}_b\}\rangle
\label{eq:finalsy2}
\\
\frac{\langle \{\hat{S}_y,\hat{S}_z\}\rangle}{N} &=& 
-\frac{1}{2N} \left(\frac{N}{2}+\langle \hat{F}_R\rangle\right) 
\langle \{\hat{N}_a-\hat{N}_b, \hat{\theta}_a-\hat{\theta}_b\}\rangle
\label{eq:finalsysz}
\eea
Note that these expressions hold independently of the approximation performed 
in the phase difference operator (that neglects the oscillating terms).

\subsection{Results of the expansion method for $\xi^2(t)$}
\label{subsec:rotemfx2}

To obtain an expression for $\xi^2$ in the double thermodynamic and Bogoliubov limit,
it remains to explicitly evaluate (\ref{eq:finalsx},\ref{eq:finalsy2},\ref{eq:finalsysz})
and to insert the result in (\ref{eq:xi2def},\ref{eq:dspm}).
The required operators have their expression given in useful form in the Appendix
\ref{app:details}, and by (\ref{eq:rephase},\ref{eq:init_diff_theta},\ref{eq:Dut}) for the phase difference operator.

We have first evaluated (\ref{eq:finalsx},\ref{eq:finalsy2},\ref{eq:finalsysz}) at time $t=0^+$, and we have checked
that one recovers the exact relations  $\langle \hat{S}_x \rangle(0^+) = N/2$, $\langle \hat{S}_y^2 \rangle(0^+) = N/4$ and $\langle \{\hat{S}_y,\hat{S}_z\}\rangle(0^+)=0$.
At finite time, squaring (\ref{eq:rephase}), one realizes that the crossed term, linear in time, has a zero
expectation value since
\be
\langle (\hat{\theta}_a-\hat{\theta}_b)(0^+) (\hat{N}_a-\hat{N}_b)\rangle = \langle (\hat{\theta}_a-\hat{\theta}_b)(0^+) \hat{\cal D}\rangle
=0
\ee
This is due to the fact that the phase difference operator at $t=0^+$ is proportional to the antihermitian part
$i\hat{y}$ of $e^{-i\hat{\theta}_a^{(0)}} \hat{b}_\oo^{(0)}$, whereas $\hat{N}_a-\hat{N}_b$ only involves the hermitian
part $\hat{x}$ and $\hat{\cal D}$ does not depend on that operator. As a consequence, the spreading of the relative
phase is purely ballistic within our Bogoliubov approximation, that it ignores the interactions among the quasi-particles 
\be
\langle (\hat{\theta}_a-\hat{\theta}_b)^2\rangle = \langle (\hat{\theta}_a-\hat{\theta}_b)^2\rangle(0^+)
+(\chi t)^2 \langle (\hat{N}_a-\hat{N}_b+\hat{\cal D})^2\rangle
\ee
(the inclusion of these interactions within a quantum
kinetic framework introduces a diffusive component in the phase spreading \cite{Sinatra:2009}).
Another simplification takes place, for similar reasons, 
\be
\label{eq:cslv}
\langle \hat{F}_I (\hat{N}_a-\hat{N}_b+\hat{\cal D})\rangle \in i \mathbb{R}
\ee
so the last term of (\ref{eq:finalsy2}) reduces to its $t=0^+$ value.
Using (\ref{eq:varsz}), and the fact that $\langle \hat{F}_R\rangle$, $\langle \hat{\cal D}^2\rangle$ and
$\langle \{\hat{S}_z,\hat{\cal D}\}\rangle$ are all $\approx N \epsilon_{\rm Bog}$, and
neglecting contributions in $\epsilon_{\rm Bog}^2$, we finally obtain
\bea
\label{eq:finalAsN}
\frac{\langle \hat{S}_y^2\rangle -\langle \hat{S}_z^2\rangle}{N} &=&
\tau^2
\left(1+ \frac{\langle \hat{\cal D}^2\rangle}{N}\right) +\frac{\langle \hat{F}_R\rangle}{N} \\
\frac{\langle \{\hat{S}_y,\hat{S}_z\}\rangle}{N} &=& \tau
\label{eq:finalBsN}
\eea
where we have introduced a dimensionless ``time"
\be
\tau(t) \equiv \frac{\rho g t}{2\hbar}  \left(1+\frac{2 \langle \hat{F}_R(t)\rangle + \langle \{\hat{S_z},\hat{\cal D}\}\rangle}{N}\right)
\label{eq:deftau}
\ee
that is slightly renormalized by a time dependent contribution of order $\epsilon_{\rm Bog}$.
The expectation values $\langle \hat{F}_R(t)\rangle$ and 
$\langle \{\hat{S_z},\hat{\cal D}\}\rangle$ appearing in (\ref{eq:deftau}) are given in the Appendix
\ref{app:details}, they are respectively uniformly bounded in time and time independent.
Note that the long-time behaviors of (\ref{eq:finalAsN},\ref{eq:finalBsN}) were expected:
As the squeezing dynamic occurs, $\hat{S}_y$ indeed grows quadratically in time while $\hat{S}_z$ stays constant. 

\begin{figure}[tb]
\begin{center}
\resizebox{0.75\columnwidth}{!}{\includegraphics{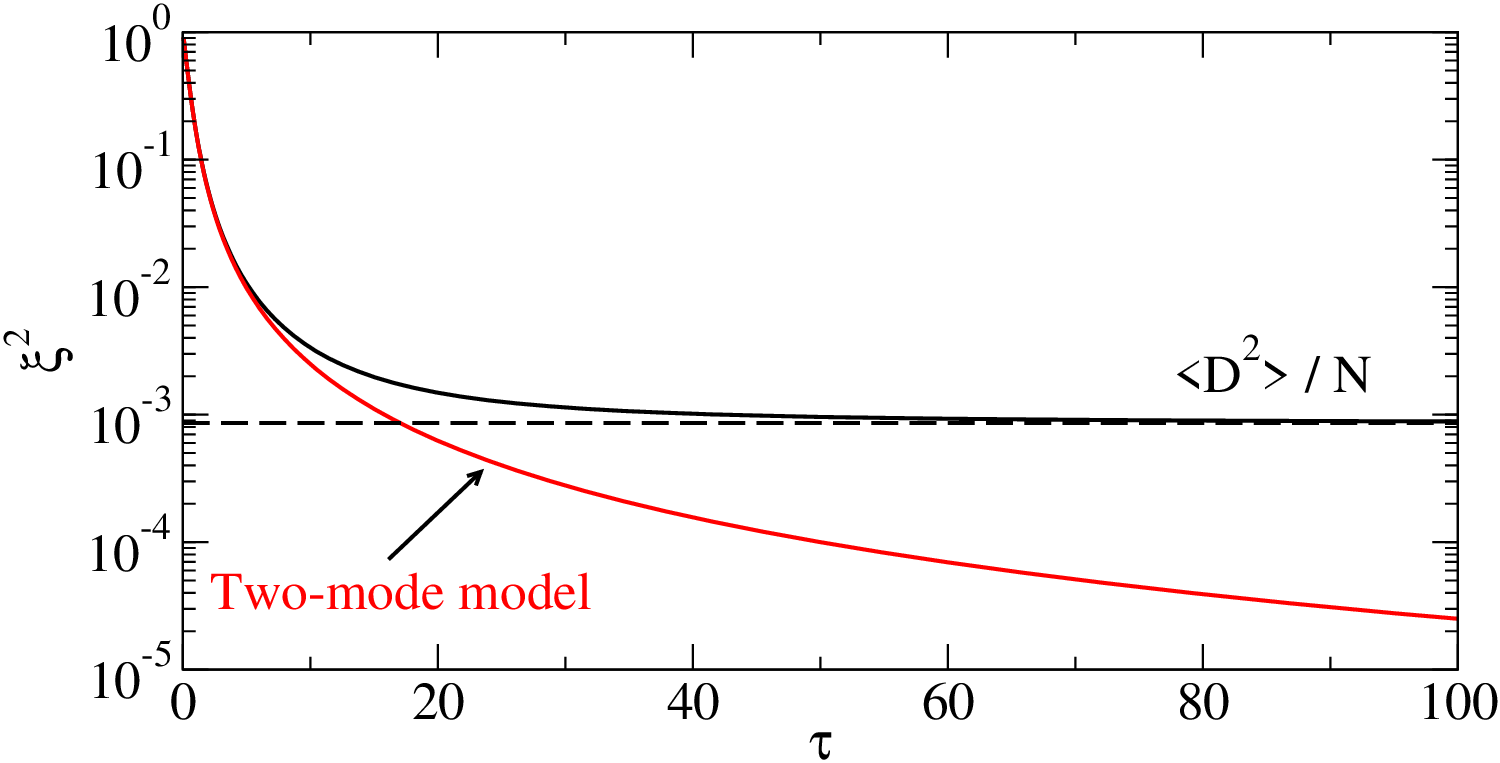} }
\caption{(Color online) Black solid line: Squeezing parameter in the thermodynamic and weakly interacting limit (\ref{eq:central}) as a function of the renormalized time $\tau$ (\ref{eq:deftau}). Parameters: $k_BT/\rho g = 1$ and $\sqrt{\rho a^3}=10^{-3}$.
Black dashed line: asymptotic value (\ref{eq:xi2min_d}). Red solid line: two-mode model prediction.  
\label{fig:xi2}}
\end{center}
\end{figure}

Finally, expanding (\ref{eq:dspm}) up to order one included in $\epsilon_{\rm Bog}$ at any fixed $\tau$ gives
the squeezing parameter as a function of time in the thermodynamic limit
(in particular for $\tau \ll N^{1/2}$):
\be
\label{eq:central}
\xi^2(t) \simeq \frac{1-4 \frac{\langle\hat{F}_R\rangle}{N}}{(\tau+\sqrt{1+\tau^2})^2} +
\frac{2\left(\frac{\langle\hat{\cal D}^2\rangle}{N} \tau^2 + \frac{\langle\hat{F}_R\rangle}{N}\right)}
{(\tau+\sqrt{1+\tau^2})\sqrt{1+\tau^2}}
\ee
As we show in Fig.\ref{fig:xi2}, the squeezing parameter $\xi^2(t)$ decreases in time
essentially as in the two-mode model 
[this is the first term in the right hand side of (\ref{eq:central}) without the ${\langle\hat{F}_R\rangle}/{N}$ term,
see e.g. equation (52) in \cite{Sinatra:Frontiers})] until a renormalized time $\tau\approx 1/\sqrt{\epsilon_{\rm Bog}}\gg 1$ is reached, when the multimode effects [the second term] start limiting the squeezing.
At such times, the contribution of $\langle\hat{\cal D}^2\rangle/N$ dominates over the one of $\langle\hat{F}_R\rangle/N$
by a factor $\approx 1/\epsilon_{\rm Bog}^2$, which shows that $\langle\hat{\cal D}^2\rangle/N$
constitutes the real actor in the process of squeezing limitation.

\subsection{Minimal squeezing and best squeezing time}
\label{subsec:results}

\begin{figure}[tb]
\begin{center}
\resizebox{0.75\columnwidth}{!}{\includegraphics{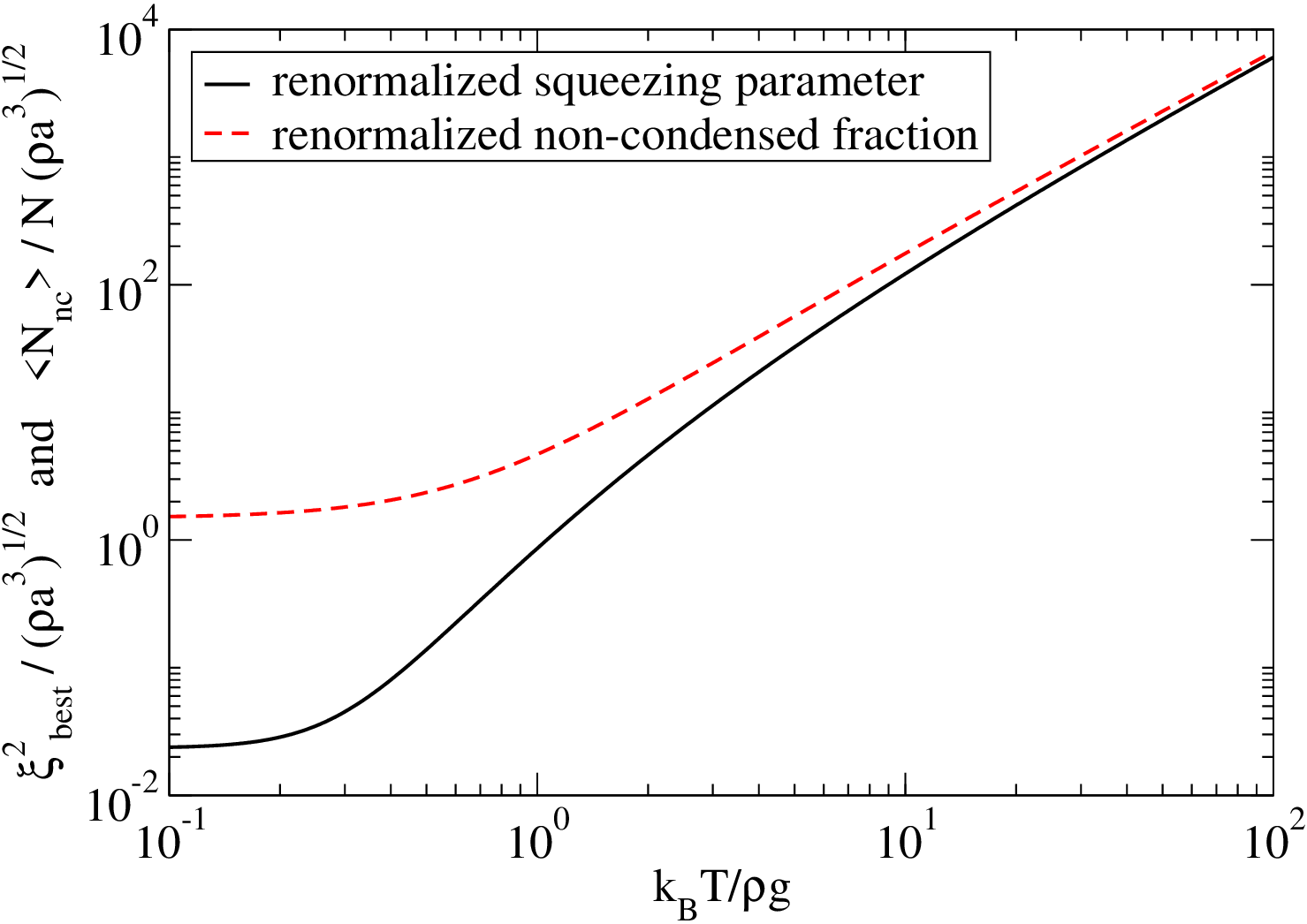}}
\caption{(Color online).\ Minimal squeezing (black solid line) $\xi_{\rm min}^2$ and non condensed fraction (red dashed line) $\langle N_{\rm nc} \rangle/N$, both divided by $\sqrt{\rho a^3}$, as a function of $k_BT/\rho g$.
\label{fig:inset_seul}}
\end{center}
\end{figure}

From the central result (\ref{eq:central}) of the previous subsection, it appears 
that the minimal squeezing $\xi_{\rm min}^2$ is reached in the thermodynamic and Bogoliubov limits 
at ``infinite" time and it is given by 
\be
\xi^2_{\rm min} = {{\langle \hat{\cal D}^2 \rangle} \over N} 
\label{eq:xi2min_d}
\ee
An explicit calculation gives:
\begin{equation}
\xi_{\rm min}^2 \!=\!\!
\int\!\! {d^3 k \over (2\pi)^3} \, {s_k^4 \over 2\rho} \left[ \Big(n_k^{(0)}+{1\over 2}\Big)\! 
\left( {  (s_k^{(0)})^2\over  s_k^4 } + {s_k^4 \over  (s_k^{(0)})^2 }  \right) -1 \right] 
\label{eq:xibest_q}
\end{equation}
where we have introduced the mean occupation numbers $n_k^{(0)}=1/[\exp(\beta\epsilon_k^{(0)})-1]$ of the Bogoliubov
quasi-particles in the initial ($t=0^-$) thermal equilibrium gas in internal state $a$.
The prediction of (\ref{eq:xibest_q}) is shown as a dashed line in Fig.\ref{fig:xibest} and as a full line in
Fig.\ref{fig:inset_seul}. In  Fig.\ref{fig:inset_seul} we show that the minimal squeezing parameter given by (\ref{eq:xibest_q}) is always lower than the non condensed fraction $\langle N_{\rm nc} \rangle/N$ where $\langle N_{\rm nc}\rangle \equiv \langle \hat{N}_{a\perp}^{(0)}\rangle$ 
is the mean number of non condensed atoms in component $a$ before the pulse: 
\be
\langle N_{\rm nc} \rangle = \sum_{\kk \neq {\bf 0}} \left[\left(U_k^{(0)}\right)^2+\left(V_k^{(0)}\right)^2\right] n_k^{(0)} +\left(V_k^{(0)}\right)^2 
\ee
Asymptotically, for $k_BT \gg \rho g$ (but always $T\ll T_c$) the minimal squeezing parameter $\xi^2_{\rm min}$ reaches the non condensed fraction.
In the opposite limit, for $k_BT / \rho g\to 0$, the squeezing tends to a constant value. 
\begin{equation}
{\xi_{\rm min}^{2 \, (T=0)} \over \sqrt{\rho a^3} }= \sqrt{{8\over \pi}} \left[ {19\over 6}\sqrt{2} -{3\over 2} {\rm ln}(\sqrt{2}+1)-\pi\right] \simeq 0.02344 \label{eq:xi2T0} 
\end{equation}
Although non zero, $\xi_{\rm min}^{2 \, (T=0)}$ is very small for practical purposes.
Indeed  $\rho a^3 < 10^{-6}$ in present squeezing experiments so that (\ref{eq:xi2T0}) predicts $\xi_{\rm min}^{2 \, (T=0)}  \aplt 2 \times 10^{-5}$.

\begin{figure}[tb]
\begin{center}
\resizebox{0.75\columnwidth}{!}{\includegraphics{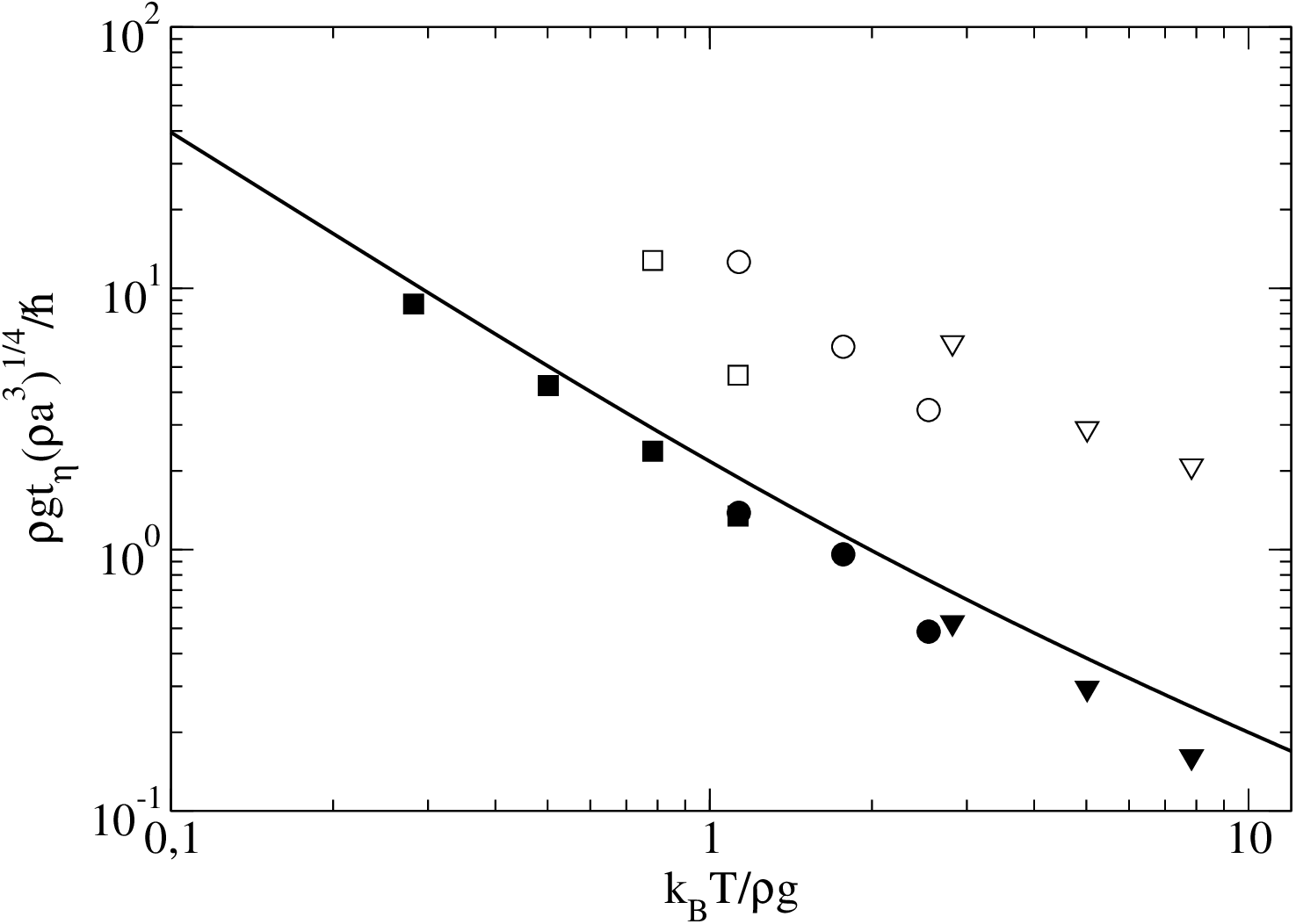} }
\caption{Universal scaling of the ``close to best" squeezing time $t_\eta$ with $\eta=0.2$. Filled symbols: classical field simulations with
$\sqrt{\rho a^3}=1.32 \times10^{-2} $ (squares), $1.94 \times10^{-3}$ (disks) and $\sqrt{\rho a^3}=4.317 \times10^{-4} $ (triangles). 
Solid line: analytical result (\ref{eq:t_eta}) using (\ref{eq:xibest}). The empty symbols (squares, circles and triangles) show the
thermalization times $t_{\rm therm}$ extracted  from the simulations as explained in \cite{Sinatra:2011}. Parameters are as in Fig.\ref{fig:xibest}.
\vspace{-0.5cm}
\label{fig:tratio}}
\end{center}
\end{figure}

The fact that the minimal squeezing is obtained at infinite time is a limitation of our Bogoliubov approach, that
neglects the interactions between the quasi-particles and effectively assumes that the corresponding collision time 
is diverging, see discussion below.
However, since the numerical squeezing curve $\xi^2(t)$ is quite flat around its minimum (see Fig.\ref{fig:xi02}), it
 suffices in practice to determine a ``close to best" squeezing time $t_\eta$ defined
 as $\xi^2(t_\eta)=(1+\eta) \xi_{\rm min}^2$, where $\eta>0$. Then, according to (\ref{eq:central}) expanded for large $\tau$
up to order $\tau^{-2}$ included, $t_\eta$ is finite and given for $\eta\ll 1$ by
\begin{equation}
\frac{\rho g}{\hbar} t_\eta \simeq {1 \over \sqrt{\eta \xi^2_{\rm min}}} \label{eq:t_eta}
\end{equation}
The ``close to best" squeezing time $t_\eta$ (\ref{eq:t_eta}) for $\eta=0.2$ is shown in Fig.\ref{fig:tratio} as a full line and compared to simulations
(filled symbols).

\begin{figure}[tb]
\begin{center}
\resizebox{0.75\columnwidth}{!}{\includegraphics{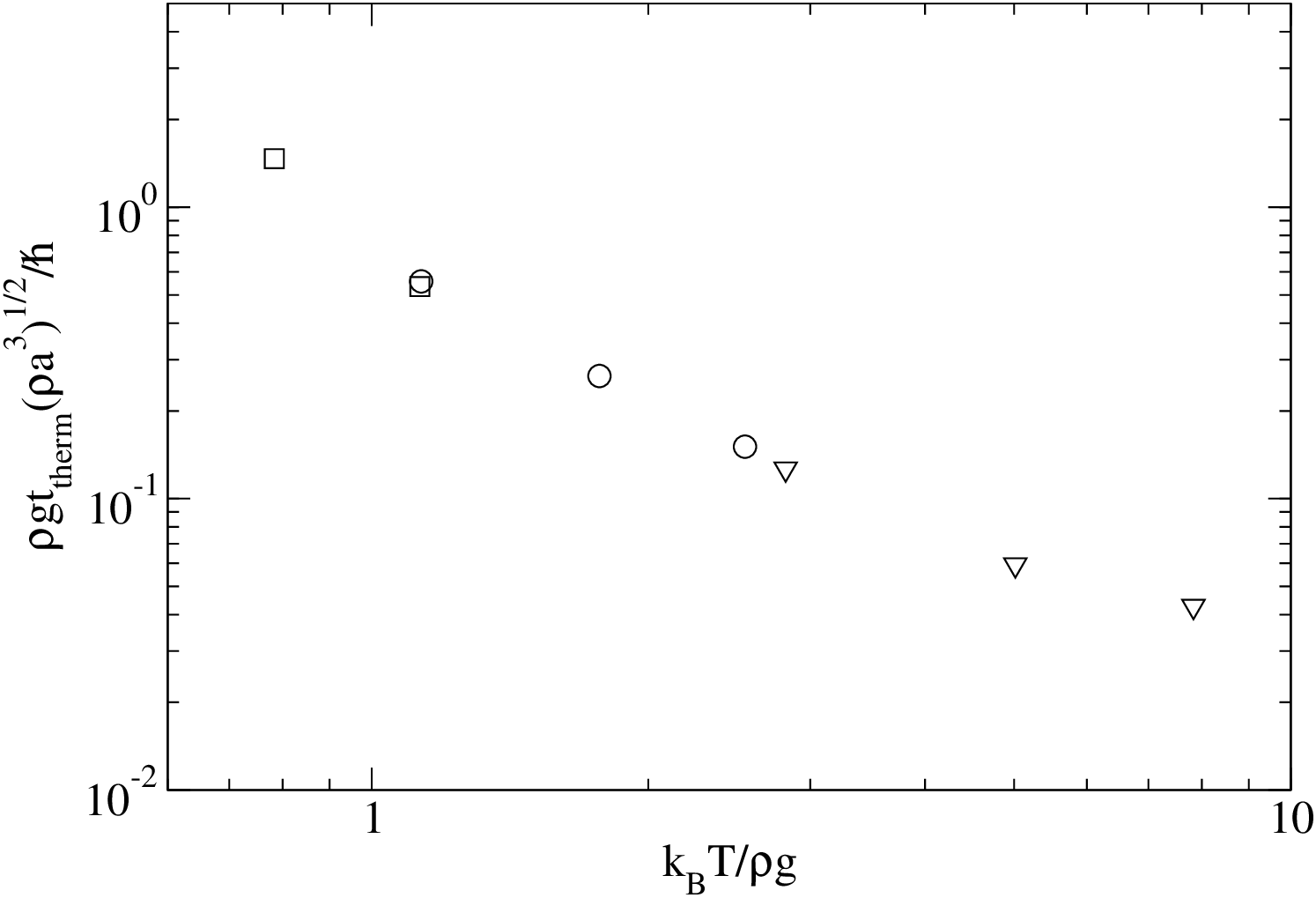} }
\caption{Scaling with $(\rho a^3)^{-1/2}$ of the thermalization times extracted from the classical field simulations.
$\sqrt{\rho a^3}=1.32 \times10^{-2} $ (squares), $1.94 \times10^{-3}$ (circles) and $\sqrt{\rho a^3}=4.317 \times10^{-4} $ (triangles). 
\vspace{-0.5cm}
\label{fig:t_th}}
\end{center}
\end{figure}

An important issue is that of {\it thermalization} that brings the system back to equilibrium after the pulse. Thermalization
is neglected in the Bogoliubov theory and in our analytics but it is included in the classical field simulations.
It is thus possible to reach $\xi^2 = (1+\eta) \xi_{\rm min}^2$ only if
$t_\eta$ given by (\ref{eq:t_eta}) is shorter than the thermalization time:
\begin{equation}
t_\eta < t_{\rm therm}  \label{eq:cond_t}
\end{equation}
We show the thermalization times, that we extract from the classical simulations as explained in \cite{Sinatra:2011}, as empty symbols in Fig.\ref{fig:tratio}.
For the points we considered they are indeed longer than the close to best squeezing times and the condition (\ref{eq:cond_t})
is satisfied.
We can also estimate $t_{\rm therm}$ from Landau-Beliaev damping rates of Bogoliubov modes  \cite{Sinatra:2008,Giorgini:1998}.
The damping rate of the mode ${\bf q}$ is 
\begin{equation}
\Gamma_q= \frac{g}{2\pi^2\hbar\xi_{\rm heal}^3} \left( \check{\Gamma}_q^L + \check{\Gamma}_q^B\right)
\label{eq:rescaled_gamma}
\end{equation}
where the healing length $\xi_{\rm heal}$ is defined by $\rho g=\hbar^2/(2m\xi_{\rm heal}^2)$
and the rescaled Landau and Beliaev damping rates $\check{\Gamma}_q^L $ and  $\check{\Gamma}_q^B$ are dimensionless functions of 
$k_B T/\rho g$ only, given e.g. in equations (A7) and (A13) of \cite{Sinatra:2008}. 
Concentrating on the pre-factor in (\ref{eq:rescaled_gamma}), 
for fixed $k_BT/\rho g$, this gives 
\be
\frac{\rho g t_{\rm therm}}{\hbar} \simeq \frac{\rho g }{\hbar \Gamma_q}= \sqrt{ \frac{\pi}{128} }
\left( \check{\Gamma}_q^L + \check{\Gamma}_q^B \right)^{-1}  \frac{1}{\sqrt{\rho a^3} } 
\ee 
The scaling with $(\rho a^3)^{-1/2}$ of the thermalization time is shown in Fig.\ref{fig:t_th}.
On the other hand, the close to best squeezing time $t_\eta$ scales as 
\be
\frac{\rho g t_{\eta}}{\hbar} \propto \frac{1}{\left( \rho a^3 \right)^{1/4}} \ \ \ \mbox{so\ that} \ \ \ 
{t_\eta \over t_{\rm therm}} \propto (\rho a^3)^{1/4}
\ee
and (\ref{eq:cond_t}) is satisfied in the weakly interacting limit.

\section{Physical interpretation}
\label{sec:interpret}

\subsection{Limit to the squeezing}

In the spin squeezing scheme with condensates that we consider, the useful quantum correlations are built through mean field 
interactions that introduces a phase shift for each atom that depends on the collective variable $N_a-N_b$. In the collective spin picture,
we can say that in a given realization of the experiment, the component $\hat{S}_y$ becomes an enlarged copy of $\hat{S}_z$ so that correlations
build up in the $S_y$-$S_z$ plane, orthogonally to mean spin. To explain this fact in a simple reasoning, we can identify $\hat{S}_y$ with
the condensate relative phase: $\hat{S}_y\simeq{N\over 2}(\hat{\theta}_a-\hat{\theta}_b)$ and look at equation (\ref{eq:rephase}) that we rewrite here
replacing $\chi$ by its expression $\chi=g/(\hbar V)$:
\be
\hat{\theta}_a-\hat{\theta}_b=(\hat{\theta}_a-\hat{\theta}_b)(0^+) - \frac{\rho g t}{\hbar N}  \left[ \hat{N}_a-\hat{N}_b  + \hat{\cal D} \right] 
\label{eq:phase2}
\ee
Initially, at $t=0$, the phase difference $(\hat{\theta}_a-\hat{\theta}_b)(0^+)$ is of order $1/\sqrt{N}$ and $N_a-N_b$ is of order
$\sqrt{N}$.  As soon as $\rho g t/\hbar \gg 1$, the time dependent term in (\ref{eq:phase2}) dominates over the initial condition. 
In the absence of the multimode  contribution to the phase difference (i.e. for $\hat{\cal D}=0$), 
$\hat{\theta}_a-\hat{\theta}_b$ and thus $\hat{S}_y$
become an enlarged copy of $\hat{S}_z=(\hat{N}_a-\hat{N}_b)/2$. 
This is the scenario in the two-mode theory. Correlations between $\hat{S}_y$ and $\hat{S}_z$
becomes perfect in the long time limit. In this case there is no limit to the squeezing and $\xi^2_{\rm min} \to 0$ when $N\to \infty$. 
On the other hand, in the presence of  $\hat{\cal D}$ this is not possible.
Looking at squeezing in the long time limit where $|\hat{S}_y| \gg |\hat{S}_z|$ and keeping only the 
leading ($0^{\rm th}$) order in $1/t$ in equation 
(\ref{eq:central}), we can write
\begin{equation}
\xi^2(t) \simeq \frac{\langle \hat{S}_y^2\rangle \langle \hat{S}_z^2 \rangle - \langle \{ \hat{S}_y , \hat{S}_z \}/2 \rangle^2}
{\langle \hat{S}_y^2 \rangle \langle \hat{S}_z^2 \rangle} \simeq {\langle \hat{\cal D}^2 \rangle \over N} =\xi_{\rm min}^2  
\end{equation}
The only contribution left in $\xi^2_{\rm min}$ is the variance of $\hat{\cal D}$ that is the part of ${\theta}_a-{\theta}_b$ that is not proportional to $N_a-N_b$. But what is the physical origin of $\hat{\cal D}$, given by (\ref{eq:calD}) ? It comes from the fact that the mean field interaction for a condensed atom 
with and another condensed atom or with an atom in an excited mode is not the same. This is particularly clear in the Hartree-Fock limit
where $V_k\to 0$ and $U_k\to 1$. In this case $\hat{\cal D}$ reduces to $\hat{N}_{a\perp}-\hat{N}_{b\perp}$
that is the non-condensed atom number difference.
In this limit we have
\be
\left(\hat{\theta}_a-\hat{\theta}_b\right)_{\rm HF} \simeq - 
\frac{\rho g t}{\hbar N}  \left[ \hat{N}_{a {\bf 0}}-\hat{N}_{b {\bf 0}} + 2\left( \hat{N}_{a\perp}-\hat{N}_{b\perp} \right)  \right] 
\ee
the factor $2$ is the Hartree-Fock factor that doubles the effective strength of the condensate-non condensate interaction with respect to
the condensate-condensate interaction.

\subsection{Squeezing of the condensate mode}
\begin{figure}[htb]
\begin{center}
\resizebox{0.75\columnwidth}{!}{\includegraphics{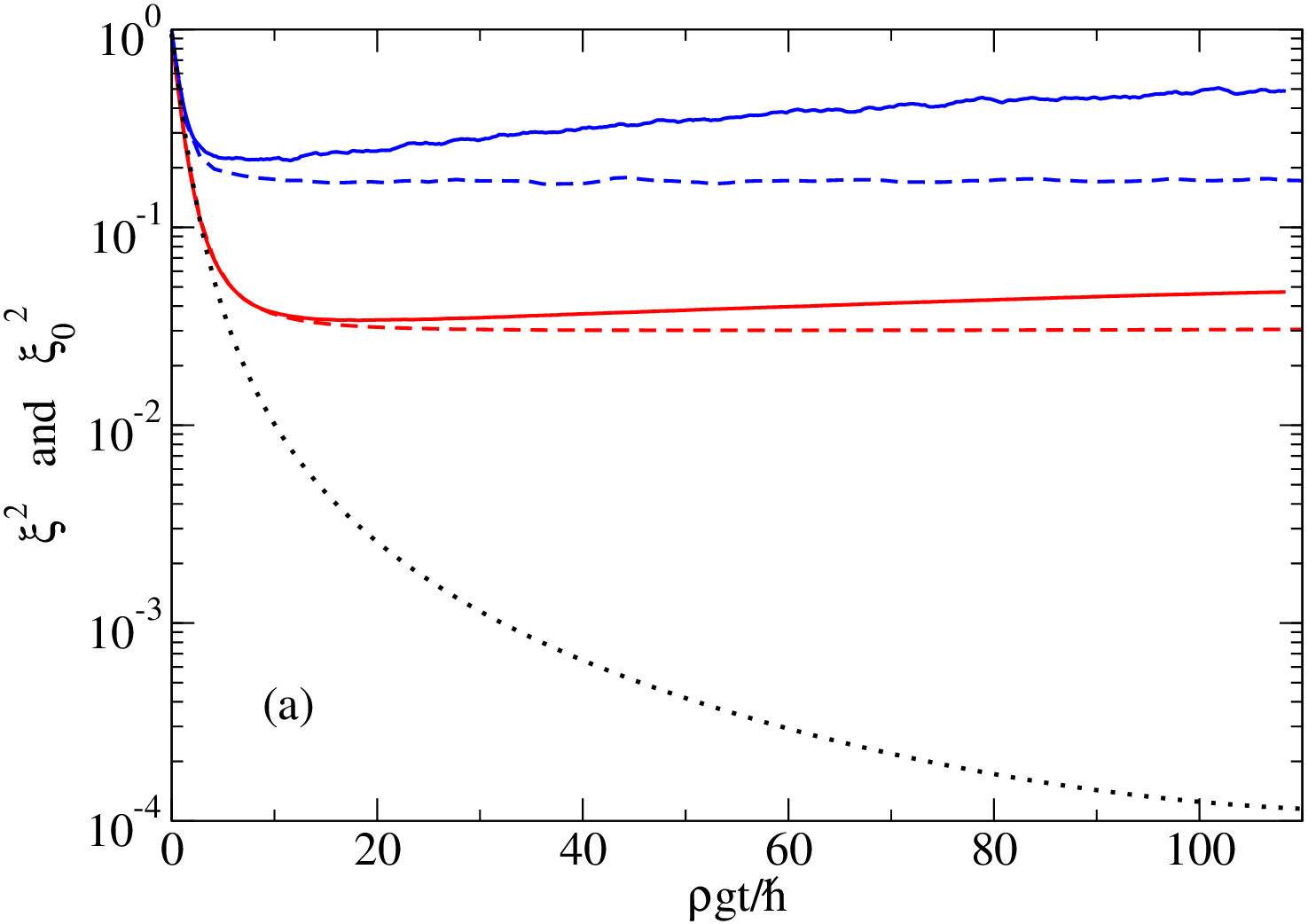} }
\resizebox{0.75\columnwidth}{!}{\includegraphics{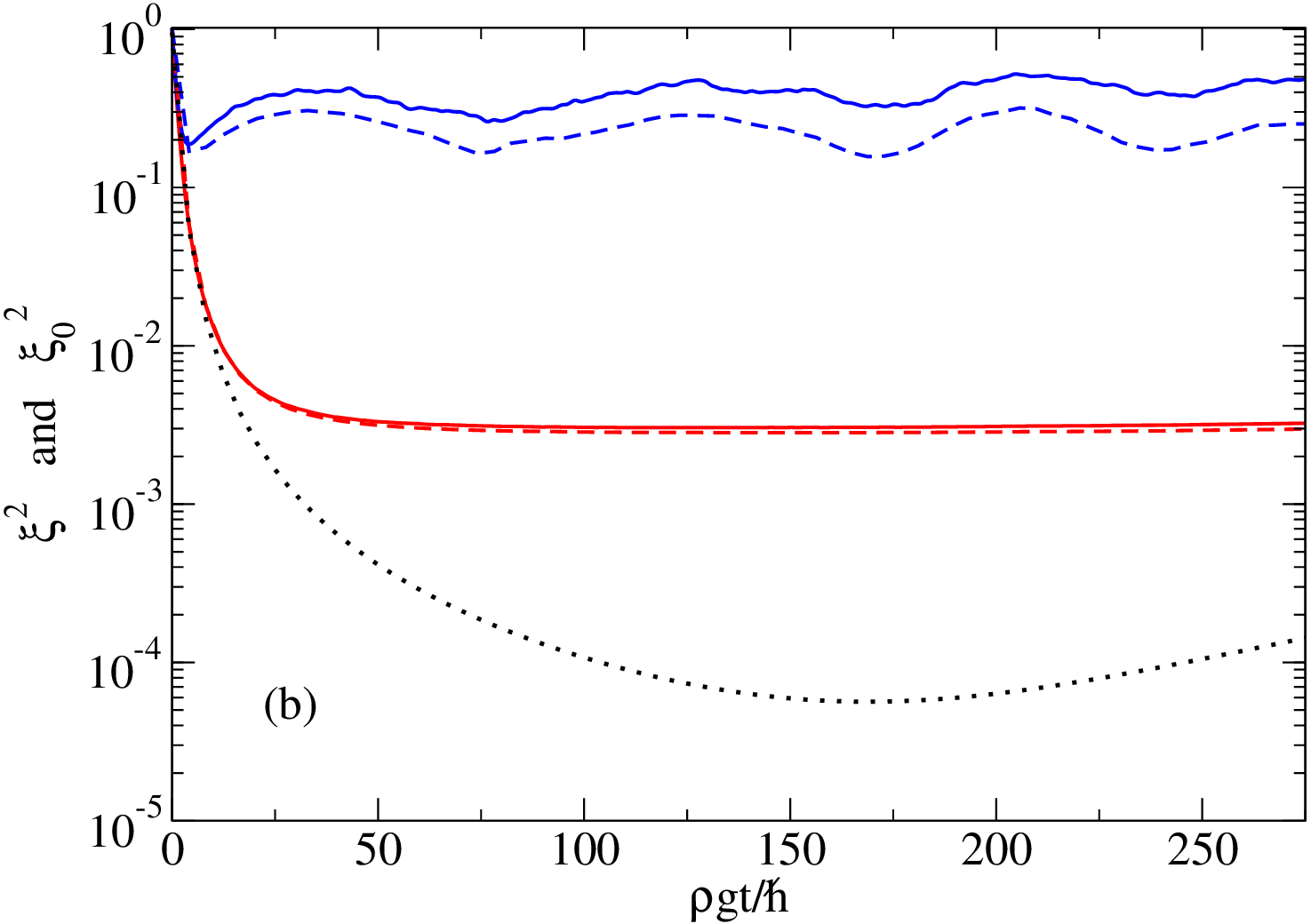} }
\caption{(Color online).\
Condensate squeezing $\xi_0^2$ (blue upper curves) and squeezing of the total field  $\xi^2$ (red middle curves) as a function of time. 
Full lines: classical field simulation. Dashed lines: Complete classical field
Bogoliubov approximation implemented numerically.
The two-mode result (black lowest dotted lines) is shown for comparison. 
Upper graph: $k_B T/\rho g=7.83$, $\langle N_{\rm nc} \rangle/N=0.05$, $N=983040$, $\rho g=693.6 \hbar^2/mV^{2/3}$,  $\sqrt{\rho a^3}=4.17\times 10^{-4}$. 
Lower graph: $k_B T/\rho g=0.5$, $\langle N_{\rm nc} \rangle/N=0.02$, $N=2733750$, $\rho g=13715.9 \hbar^2/mV^{2/3}$, $\sqrt{\rho a^3}=1.32\times 10^{-2}$. 
\label{fig:xi02}}
\end{center}
\end{figure}

In Fig.\ref{fig:xi02}, for two temperatures: $k_BT\gg\rho g$ and $k_BT<\rho g$,
we compare the squeezing of the total field $\xi^2$, that we have been considering so far, with the squeezing
of the condensate mode $\xi_0^2$, constructed with a spin operator involving the condensate mode only: 
$\hat{S}_{{\bf 0} x}+i \hat{S}_{{\bf 0} y}=\hat{a}_{\bf 0}^\dagger \hat{b}_{\bf 0}$ and $\hat{S}_{{\bf 0} z} = \hat{N}_{a {\bf 0}}-\hat{N}_{b {\bf 0}}$:
\begin{equation}
\xi_0^2=\frac{\langle N_{a{\bf 0}}\rangle \Delta S_{{\bf 0} \perp,\mbox{\small min}}^2}{|\langle {\bf S_{{\bf 0}}}\, 
\rangle|^2 } \, , \label{eq:xi20def}
\end{equation}
This is the squeezing that would be obtained by ``selecting" only the condensed particles for the squeezing measurement.
Besides the classical field simulations, in Fig.\ref{fig:xi02}a and Fig.\ref{fig:xi02}b we also show as
dashed curves results obtained in the Bogoliubov approximation \footnote{
Before the pulse, we start with a thermalized field sampling (\ref{eq:rho_can_cl}). After the pulse,
we evolve the condensate phase with the classical equivalent of (\ref{eq:thetadotsigma}), also performing in that equation
the approximation of neglecting the oscillating terms, and the Bogoliubov amplitudes with (\ref{eq:ckt}).
The condensate atom numbers are obtained by the classical equivalent of (\ref{eq:conservation}).}. Clearly $\xi^2 \ll \xi_0^2$ in both graphs. Particularly striking is the case in Fig.\ref{fig:xi02}b where $\xi_{0 {\rm min}}^2/\xi_{\rm min}^2\simeq 60$
while the non condensed fraction is only $\langle N_{\rm nc} \rangle/N=0.02$.
We explain here why this is the case. 
In order to have condensate squeezing we need to build up correlations between $\hat{S}_{{\bf 0}y}$, that is still proportional
to $\hat{\theta}_a-\hat{\theta}_b$, and $\hat{S}_{{\bf 0} z}$. According to (\ref{eq:phase2}),  at long times
$\hat{\theta}_a-\hat{\theta}_b$ differs from $\hat{S}_{{\bf 0} z}$ by the quantity $\hat{N}_{a\perp}-\hat{N}_{b\perp} + \hat{\cal D}$
that prevents the correlations to become perfect at long times. Indeed, we find at long times and for a large system that
\be
\xi_{0}^2(t) \sim \frac{{\mbox{Var}}\left[ \left(\hat{N}_{a\perp}-\hat{N}_{b\perp} \right) + \hat{\cal D} \right]}{N}
\label{eq:xi02}
\ee
The evaluation of the minimal achievable values of $\xi_0^2$ from (\ref{eq:xi02}) is more involved than for $\xi^2$
because the quantity $\hat{N}_{a\perp}-\hat{N}_{b\perp}$, contrarily to $\hat{\cal D}$,
is not a constant of motion. A detailed discussion is beyond the scope of this paper, but one can give simple
reasons explaing why the minimal value of $\xi_0^2$ is numerically found to be much larger than the one of $\xi^2$.
Let us first forget about the time dependence of $\hat{N}_{a\perp}-\hat{N}_{b\perp}$ and evaluate it at time $t=0^+$.
It is found that the variance of $\hat{N}_{a\perp}-\hat{N}_{b\perp}$ at $t=0^+$ is simply $\langle N_{a\perp}^{(0)}\rangle$,
that is the mean number $\langle N_{\rm nc}\rangle$ of non-condensed particles before the pulse.
In the Hartree-Fock limit $k_B T \gg \rho g$, $\hat{\cal D}$ reduces to $\hat{N}_{a\perp}-\hat{N}_{b\perp}$. One then 
expects that the minimal $\xi_0^2$ is four times the non-condensed fraction, 
that is four times larger
than $\xi_{\rm min}^2$. In the low temperature regime $k_B T \ll \rho g$, $\langle \hat{\cal D}^2\rangle \ll \langle N_{\rm nc}\rangle$,
so the ratio of condensate to total field squeezing is expected to be even larger.

Let us now take into account the time dependence of $\hat{N}_{a\perp}-\hat{N}_{b\perp}$. This makes the situation
even worse for the condensate mode squeezing:
Whereas the operator $\hat{\cal D}$ has normal fluctuations with a variance scaling as the volume $V$ of the system, it is found
that $\hat{N}_{a\perp}-\hat{N}_{b\perp}$ has anomalous fluctuations at long times.
If one replaces discrete sums over ${\bf k}$ by integrals in the expression (\ref{eq:vardnp}) of the variance of
$\hat{N}_{a\perp}-\hat{N}_{b\perp}$, as usual in the thermodynamic limit, one finds converging integrals but the variance 
diverges linearly in the long time limit:
\be
\left[\frac{\mbox{Var}\, (\hat{N}_{a\perp}-\hat{N}_{b\perp})}{N}\right]_{\rm therm.lim.}\!\!\!\! \sim 
\frac{3}{2} (2\pi\rho a^3)^{1/2} \frac{k_B T t}{\hbar} 
\label{eq:var_nperpdiff}
\ee
This result obtained within the Bogoliubov theory of course fails
for times larger than the thermalisation time.
In practice, it is more physical to consider a finite size system. One then finds that, in the long time limit, 
the variance is dominated by the contribution of the low-$k$ terms in the sum
over ${\bf k}$ in (\ref{eq:vardnp}): The variance of $\hat{N}_{a\perp}-\hat{N}_{b\perp}$ grows from its extensive $t=0^+$ value 
$\langle \hat{N}_{a \perp}^{(0)} \rangle$ to a super-extensive value 
$\simeq 10 (k_BT/\rho g)(mcL/2\pi\hbar)^{4}$ \footnote{The numerical
coefficient in this formula is given for a cubic quantization box
with periodic boundary conditions, see discussion in section 7.8
of \cite{LesHouches:1999}.}
in a time scaling as $L/c$ where $L=V^{1/3}$ is the system size 
and $c=\sqrt{\rho g/m}$ is the initial sound velocity. After this time, the variance oscillates around the super-extensive value with a period again scaling as $L/c$, given by the fundamental Bogoliubov modes in the box.
This explains the oscillations of $\xi_0^{2}$ and its temporal mean value
in Fig.\ref{fig:xi02}b. In Fig.\ref{fig:xi02}a the oscillations are less visible as the
anomalous contribution to the variance is smaller than its $t=0^+$ value $\langle \hat{N}_{a \perp}^{(0)} \rangle$.

\section{Conclusion}
\label{sec:conclusions}
We have shown that, in a multimode theory, the spin squeezing that can be obtained dynamically using interactions in condensates is finite
in the thermodynamic limit. This is contrary to the results of the currently used two-mode theory that predicts an infinite metrology gain for
$N\to\infty$. Using a convenient reformulation of the Bogoliubov theory, we could calculate the temperature and interactions dependent limit
of the spin squeezing parameter analytically for a spatially homogeneous system. 
We performed non perturbative classical field simulations to test our analytical results including interactions
among Bogoliubov modes and thermalization that are neglected in the perturbative treatment. 

At temperatures $k_B T \ll \rho g$ the limit that we find for the squeezing parameter optimized over time, $\xi_{\rm min}^2$, is very small and in 
particular much smaller than what is currently measured in present experiments. Nevertheless it represent the fundamental limit of this
squeezing scheme and we hope that the temperature dependent  limitation to spin squeezing will be soon within reach of experiments.
We explained that the physical origin to this limit of the squeezing lies in the difference of mean field interactions between condensate-condensate
and condensate-non condensed atoms. 

\bigskip
A.S. acknowledges useful discussions with J. Reichel and J. Est\`eve. 
E.W. acknowledges support from CNRS and Polish GRF: N N202 128539.

\appendix

\section{Some useful relations}
\label{app:useful}
Here we collect useful commutation relations and recall the
expression of the Bogoliubov quasi-particle annihilation operators $\hat{c}_{\sigma\kk}$ after the mixing pulse
in terms of the atomic annihilation and creation operators.

\noindent{\bf Commutation relations:} In our lattice model,
the field operators obey discrete bosonic commutation relations: 
\be
[{\hat{\psi}}_\sigma(\rr),{\hat{\psi}}_{\sigma'}^\dagger(\rr')] = \frac{\delta_{\rr,\rr'} \, \delta_{\sigma,\sigma'}}{dV}  
\ \ \ \ \forall \sigma,\sigma'=a,b
\label{eq:comm_psi}
\ee
The hermitian condensate number and phase operators obey, for $\sigma,\sigma'=a,b$:
\be
[\hat{N}_{\sigma {\bf 0}},\hat{\theta}_{\sigma'}]=i \delta_{\sigma,\sigma'}
\label{eq:comm_nb_ph}
\ee
The fields of the non-condensed modes, orthogonal to the condensate wavefunction, obey
\be
[{\hat{\psi}}_{\sigma \perp}(\rr), {\hat{\psi}}_{{\sigma'} \perp}^\dagger(\rr')] = \frac{\delta_{\sigma,\sigma'}\delta_{\rr,\rr'}}{dV} 
-\frac{\delta_{\sigma,\sigma'}}{V}
\label{eq:commpsi_perp}
\ee 
The non-condensed fields do not commute with the the total atom number but they commute with all the condensate operators:
\be
[\hat{N}_{\sigma {\bf 0}},{\hat{\psi}}_{{\sigma'} \perp}(\rr)]=[\hat{\theta}_\sigma,{\hat{\psi}}_{\sigma' \perp}(\rr)]=0
\ee
The number-conserving operators $\hat{\Lambda}_\sigma$ obey the same commutation relations (\ref{eq:commpsi_perp})
as the non-condensed fields, e.\ g.
\be
[{\hat{\Lambda}_\sigma}(\rr), {\hat{\Lambda}_\sigma}^\dagger(\rr')] = \frac{\delta_{\rr,\rr'}}{dV} -\frac{1}{V} \ \ \ \ 
\forall \sigma=a,b
\label{eq:commLambda2}
\ee 
but, contrarily to them, they commute with the total atom number operators in each component:
\be
\left[ {\hat{\Lambda}_\sigma}(\rr), {\hat{N}_{\sigma'}} \right] =0 \ \ \ \ \ \forall \sigma,\sigma'=a,b
\label{eq:commLambda}
\ee 
Their commutation relations with the condensate operators are, for $\sigma,\sigma'=a,b$:
\be
\left[ {\hat{\Lambda}_\sigma}(\rr), {\hat{\theta}_{\sigma'}} \right] =0\;, \hspace{5mm} 
\left[ {\hat{\Lambda}_\sigma}(\rr),\hat{N}_{\sigma' {\bf 0}} 
\right] =-{\hat{\Lambda}_\sigma}(\rr)\, \delta_{\sigma,\sigma'}
\ee
Finally, from the relation $e^{-i\hat{\theta}_a} f(\hat{N}_{a\oo})\, e^{i\hat{\theta}_a}
=f(\hat{N}_{a\oo}-1)$ resulting from (\ref{eq:comm_nb_ph}), for a generic function $f$,
we have in the large $N$, and thus large $\hat{N}_{a {\bf 0}}$ limit:
\begin{equation}
\left[\sqrt{\hat{N}_{a {\bf 0}}},e^{i{\hat{\theta}_a}}\right] = - e^{i \hat{\theta}_a} 
\frac{1}{2\sqrt{\hat{N}_{a {\bf 0}}} } \label{eq:comm_sqrt} + O\left(\frac{1}{N^{3/2}}\right)
\end{equation}

\noindent{\bf Bogoliubov transformations:}
By projecting (\ref{eq:Lambdat}) over the plane waves, we obtain after the pulse ($t>0$):
\begin{eqnarray}
{\hat{a}_{\bf k}}&=&e^{i {\hat{\theta}}_a} \left( \hat{c}_{a \bf k}\,  U_k + \hat{c}_{a -{\bf k}}^\dagger  V_k \right) \label{eq:invBogca} \\
{\hat{b}_{\bf k}}&=&e^{i {\hat{\theta}}_b} \left( \hat{c}_{b \bf k}\,  U_k + \hat{c}_{b -{\bf k}}^\dagger  V_k \right)  \label{eq:invBogcb}
\end{eqnarray}
where $\hat{a}_\kk$ and $\hat{b}_\kk$ annihilate an atom with internal state $a,b$ and wave vector $\kk$.  
The inverse relations are useful:
\begin{eqnarray}
\hat{c}_{a{\bf k}}&=&e^{-i {\hat{\theta}}_a} \hat{a}_{\bf k} \, U_k - \hat{a}_{-{\bf k}}^\dagger e^{i {\hat{\theta}}_a} V_k \label{eq:Bogca} \\
\hat{c}_{b{\bf k}}&=&e^{-i {\hat{\theta}}_b} \hat{b}_{\bf k} \, U_k -  \hat{a}_{-{\bf k}}^\dagger e^{i {\hat{\theta}}_b}  V_k \label{eq:Bogcb} 
\end{eqnarray}
with $U_k$, $V_k$ given by (\ref{eq:sk}). Note that it is often convenient to express $U_k$ and $V_k$ in terms of $s_k$, so that
for example
\be
U_k^2+V_k^2=\frac{1}{2}(s_k^2+s_k^{-2})\ \mbox{and}\ 2U_kV_k=\frac{1}{2}(s_k^2-s_k^{-2})
\ee

\section{After-pulse values of the condensate phases and quasi-particle annihilation operators}
\label{app:initial_phase}
To determine the condensate phase operator for the internal state $\sigma$ at time $t=0^+$ in terms of operators
at $t=0^-$, 
we use the fact resulting from (\ref{eq:pulse1},\ref{eq:pulse2}) that $a_\oo(0^+)=[a_\oo^{(0)}-b_\oo^{(0)}]/\sqrt{2}$
and $b_\oo(0^+)=[a_\oo^{(0)}+b_\oo^{(0)}]/\sqrt{2}$.
Then the definition (\ref{eq:mpra}) leads to
\be
e^{i\hat{\theta}_\sigma(0^+)}= e^{i\hat{\theta}_a^{(0)}} \left(\sqrt{\hat{N}_{a{\bf 0}}^{(0)}}\mp\tilde{b}_\oo^{(0)}\right)
\left| \sqrt{\hat{N}_{a{\bf 0}}^{(0)}} \mp  \tilde{b}_\oo^{(0)} \right|^{-1}
\ee
where the upper sign ($-$) is for $\sigma=a$ and the lower sign ($+$) for $\sigma=b$,
the modulus operator of an operator $\hat{X}$ is
\be
\label{eq:modulus_operator}
|\hat{X}| \equiv \left(\hat{X}^\dagger \hat{X}\right)^{1/2}
\ee
and we have introduced the number conserving operator 
$\tilde{b}_{\bf 0}^{(0)}=e^{-i\hat{\theta}_a^{(0)}}\hat{b}_{\bf 0}^{(0)}$. By expanding in the large $\hat{N}_{a{\bf 0}}^{(0)}$ limit, 
we obtain :
\begin{equation}
\hat{\theta}_\sigma(0^+)-\hat{\theta}_a^{(0)}=\mbox{Herm}\,\left[\mp \hat{y}\left(\hat{N}_{a{\bf 0}}^{(0)}\right)^{-1/2}   -\frac{1}{2} \{\hat{x},\hat{y}\} \left(\hat{N}_{a{\bf 0}}^{(0)}\right)^{-1} 
+ O\left({1\over N^{3/2}}\right) \right]
\label{eq:inith}
\end{equation}
We have introduced the decomposition 
$\tilde{b}_{\bf 0}^{(0)}=\hat{x}+i\hat{y}$, where $\hat{x}$ and $\hat{y}$ are hermitian
operators, the usual notation $\{\ ,\ \}$ for the anticommutator and the notation $\mbox{Herm}\, \hat{X}=(\hat{X}+\hat{X}^\dagger)/2$ for the hermitian part of
an operator $\hat{X}$. This leads to (\ref{eq:init_diff_theta}).

 To determine the quasi-particle annihilation operators $\hat{c}_{\sigma\kk}$ at time $t=0^+$ in terms of operators at time $t=0^-$,
we use the relations (\ref{eq:Bogca},\ref{eq:Bogcb}) to express them in terms of the atomic creation and annihilation
operators at time $t=0^+$, that are in turn expressed in terms of their values at $0^-$ thanks to 
(\ref{eq:pulse1},\ref{eq:pulse2}). One also needs the expansion $\exp[i\hat{\theta}_\sigma(0^+)]=\exp[i\hat{\theta}_a^{(0)}]
[1+i(\hat{\theta}_\sigma(0^+)-\hat{\theta}_a^{(0)}) +O(1/N)]$ deduced from (\ref{eq:inith}) and the Hausdorf formula.
With the short-hand notations $\tilde{a}_\kk^{(0)}=\exp(-i\hat{\theta}_a^{(0)}) \hat{a}_\kk^{(0)}$
and $\tilde{b}_\kk^{(0)}=\exp(-i\hat{\theta}_a^{(0)}) \hat{b}_\kk^{(0)}$: 
\begin{multline}
\hat{c}_{\sigma\kk}(0^+)= \frac{U_k \tilde{a}_\kk^{(0)} - V_k \tilde{a}_{-\kk}^{(0)\dagger}}{\sqrt{2}}\mp 
\frac{U_k \tilde{b}_\kk^{(0)} -  V_k \tilde{b}_{-\kk}^{(0)\dagger}}{\sqrt{2}} 
\\
  -i\left[\hat{\theta}_\sigma(0^+) -\hat{\theta}_a^{(0)}\right]
\left[\frac{U_k \tilde{a}_\kk^{(0)} + V_k \tilde{a}_{-\kk}^{(0)\dagger}}{\sqrt{2}} 
\mp \frac{U_k \tilde{b}_\kk^{(0)} + V_k \tilde{b}_{-\kk}^{(0)\dagger}}{\sqrt{2}}
\right]
+O\left(\frac{1}{N}\right)
\label{eq:cksil}
\end{multline}
where the upper, $-$ sign is for $\sigma=a$ and the lower, $+$ sign if for $\sigma=b$.
One then introduces the operators
\be
\label{eq:def0AB}
\hat{A}_\kk=U_k \tilde{a}_\kk^{(0)}\!\! -\!\! V_k \tilde{a}_{-\kk}^{(0)\dagger} \ \mbox{and}\ \hat{B}_\kk=U_k \tilde{b}_\kk^{(0)}\!\! - \!\! V_k \tilde{b}_{-\kk}^{(0)\dagger}
\ee
This directly gives (\ref{eq:defB}).
Expressing $\tilde{a}_\kk^{(0)}$ and $\tilde{a}_{-\kk}^{(0)\dagger}$ 
in terms of the pre-pulse  quasi-particle
annihilation and creation operators thanks to the $t=0^-$ equivalent of (\ref{eq:invBogca}), gives (\ref{eq:defA}). Restricting
the accuracy of (\ref{eq:cksil}) to $O(1)$ included, gives (\ref{eq:initczp}).

\section{Correlations of $\hat{A}_{\kk}$ and $\hat{B}_{\kk}$}
\label{app:GammaDelta}

Some useful properties of the operators defined in (\ref{eq:defA},\ref{eq:defB}) [or equivalently
in (\ref{eq:def0AB})] are given here.
The commutation relations of these operators are bosonic.
This means that the only non-zero commutators (considered among all possible values of $\kk$) are
\be
\left[ \hat{A}_{\kk} , \hat{A}_{\kk}^\dagger \right] = \left[ \hat{B}_{\kk} , \hat{B}_{\kk}^\dagger \right] = 1
\ee
To calculate averages of products of $\hat{A}$ and $\hat{B}$ operators (here at equal times) one can use the Wick theorem
as the initial density operator for $\{c_{a \kk}^{(0)}\}$ and $\{b_{\kk}^{(0)}\}$ is a Gaussian.
All the non-zero correlations involving the $\hat{A}_{\kk}$ can be deduced from
\bea
\label{eq:mada}
\!\!\!\!\!\!\!\langle \hat{A}_{\kk}^\dagger \hat{A}_{\kk} \rangle &=&  {1\over 2} \left( n_{\kk}^{(0)} 
+{1\over 2}\right) \left[{s_k^{(0)2}\over s_k^2} +{s_k^2\over s_k^{(0)2}}\right]\! -\! {1\over 2} \\
\!\!\!\!\!\!\!\langle \hat{A}_{\kk} \hat{A}_{-\kk} \rangle &=& {1\over 2} \left( n_{\kk}^{(0)} +{1\over 2 }\right)
\left[ {s_k^{(0)2} \over s_k^2} -{s_k^2 \over s_k^{(0)2}}\right]
\eea
with $s_k^{(0)}$ and $s_k$ defined by (\ref{eq:sk0},\ref{eq:sk}).
All the non-zero correlations involving the $\hat{B}_{\kk}$ can be deduced from
\bea
\langle \hat{B}_{\kk}^\dagger \hat{B}_{\kk} \rangle &=& V_k^2 = {1\over 4} \left( s_k^2+ {1 \over s_k^2}\right) -{1\over 2} \\
\langle \hat{B}_{\kk} \hat{B}_{-\kk} \rangle &=& - U_k V_k = - {1\over 4} \left( s_k^2 - {1 \over s_k^2}\right)
\eea
All the crossed second moments, for example of the form $\langle \hat{A} \hat{B}\rangle$ or 
$\langle \hat{A}^\dagger \hat{B}\rangle$,
are zero.

\section{Double expansion of some operators}
\label{app:details}

As explained in the main text, 
to have a vanishing error on the squeezing parameter $\xi^2$ in the
thermodynamic limit, it suffices to determine the operators $\hat{S}_y$ 
and $\hat{S}_z$ up to $\approx N^{1/2}$ included.

\noindent{\bf Case of $\hat{S}_z$:} 
The operator $\hat{S}_z=(\hat{N}_a-\hat{N}_b)/2$ is a constant of motion,
it can be evaluated at $t=0^+$, and related with (\ref{eq:pulse1},\ref{eq:pulse2})
to the fields at $t=0^-$.
Then one uses the modulus-phase representation for the condensate
operator in $a$  and one introduces the number-conserving fields
$\hat{\Lambda}_a^{(0)}$ and $e^{-i\hat{\theta}_a^{(0)}}\psi_{b\perp}^{(0)}$
for the non-condensed modes, whose Fourier components can be expressed
in terms of the operators $\hat{A}_\kk$ and $\hat{B}_\kk$ through
(\ref{eq:def0AB}). The only approximation is then to neglect the commutator
of $\hat{y}$ with $(\hat{N}_{a\oo}^{(0)})^{1/2}$, 
which is $O(1/N^{3/2})$, to obtain
\begin{multline}
\label{eq:namnb}
\hat{N}_a-\hat{N}_b \simeq -\left\{\hat{x},\sqrt{\hat{N}_{a\oo}^{(0)}}
\right\}
-\sum_{\kk \neq {\bf 0}} \left[(U_k^2+V_k^2) 
 (\hat{A}_{\bf k}^\dagger\hat{B}_{\bf k}+\hat{A}_{\bf k}\hat{B}_{\bf k}^\dagger) \right. \\ \left.
+2 U_kV_k (\hat{A}_{\bf k}^\dagger\hat{B}_{-\bf k}^\dagger+\hat{A}_{\bf k}\hat{B}_{-\bf k}) \right]
\end{multline}
We recall than $\hat{x}$ and $\hat{y}$ are defined below
(\ref{eq:inith}). The operator $\hat{S}_z$ has a zero expectation
value, and a variance exactly equal to $N/4$, as already found by a more direct
method in (\ref{eq:varsz}), so we reach the estimate
\be
\hat{S}_z \approx N^{1/2}
\ee

With more lengthy calculations, 
we now deduce $\hat{S}_y$ from the antihermitian part of $\hat{S}_+$ written in the
form (\ref{eq:defSplus}) and we then obtain $\hat{S}_y^2$ 
and $\{\hat{S}_y,\hat{S}_z\}$.

\noindent {\bf The phase difference:} We first evaluate the scaling of the phase difference $\hat{\theta}_a-\hat{\theta}_b$
in the thermodynamic limit from the writing (\ref{eq:rephase}). The contribution of
the phase difference at time $t=0^+$ scales as $1/N^{1/2}$ according to (\ref{eq:init_diff_theta}).
The contribution proportional to $\hat{N}_a-\hat{N}_b$ scales in the same way, since
the total number difference $\approx N^{1/2}$ for the binomial distribution after the $\pi/2$ pulse.
The same conclusion holds for the contribution proportional to $\hat{\cal D}$,
see (\ref{eq:odg_D}). We reach the important conclusion that, for a finite time $t$ in the thermodynamic limit,
\be
\hat{\theta}_a-\hat{\theta}_b \approx \frac{1}{N^{1/2}}
\label{eq:estim_difftheta}
\ee
As $\frac{N}{2}+\hat{F}$ is $O(N)$, it suffices to expand the exponential in (\ref{eq:defSplus})
to first order included in the phase difference to obtain
\be
\hat{S}_+=\left[ 1-i(\hat{\theta}_a-\hat{\theta}_b)+O\left({1\over N}\right)\right] \left({N\over2}+\hat{F}_R+i \hat{F}_I\right)
\label{eq:spd}
\ee
where we have split $\hat{F}=\hat{F}_R+i\hat{F}_I$ in terms of the hermitian operators $\hat{F}_R$
and $\hat{F}_I$.

\noindent{\bf The antihermitian part of $\hat{F}$:} 
The operator $\hat{F}_I$ directly contributes to the antihermitian part of $\hat{S}_+$, so it has to be evaluated
up to $\approx N^{1/2}$ included. Its exact expression is
\be
\hat{F}_I = \frac{1}{2i} \sum_{\rr} dV \left(\hat{\Lambda}_a^\dagger \hat{\Lambda}_b - \hat{\Lambda}_b^\dagger \hat{\Lambda}_a\right)
=\frac{1}{2i}\sum_{\kk \neq {\bf 0}} \left(\hat{c}_{a \kk}^\dagger \hat{c}_{b \kk} -\mbox{h.c.}\right)
\label{eq:defFI}
\ee
This corresponds to a complex scalar product between the bicomponent fields $(\hat{\Lambda_a}, \hat{\Lambda}_a^\dagger)$ 
and $(\hat{\Lambda_b}, \hat{\Lambda}_b^\dagger)$. The Bogoliubov equations of motion for spin state $\sigma$
conserve this scalar product \cite{Castin:1998}. Due to the $a-b$ symmetry, the coefficient of Bogoliubov equations of motion
are the same for the two internal states, and $\hat{F}_I$ is a constant of motion within Bogoliubov
theory. We can thus evaluate it at time $0^+$, taking into account the corrections to $\hat{c}_{\kk\sigma}(0^+)$
due to the small condensate phase change induced by the pulse, as in (\ref{eq:cksil}):
\begin{equation}
\hat{F}_I  \simeq \frac{1}{2i} \sum_{\kk \neq 0} ( \hat{A}_{\kk}^\dagger\hat{B}_{\kk}-\hat{B}_{\kk}^\dagger\hat{A}_{\kk} )
-\frac{1}{2}\left\{ \hat{y}, \frac{\hat{N}_{a \perp}^{(0)}-\hat{N}_{b \perp}^{(0)}}{\sqrt{\hat{N}_{a{\bf 0}}^{(0)}}} 
\right\}\label{eq:deltaFI}
\end{equation}
where the operator $\hat{y}$ is defined below (\ref{eq:inith}).
This correction involving $\hat{y}$ is important to ensure that $\langle S_y^2(0^+)\rangle =N/4$ as it should be.
The operator $\hat{F}_I$ has a zero expectation value, this is why the same phenomenon
as for the operator $\hat{\cal D}$ occurs. Calculating its variance, which is dominated
by the contribution of the sum over $\kk$ in (\ref{eq:defFI}),
\be
\langle \hat{F}_I^2\rangle = 
\frac{\langle \hat{N}_{a\perp}^{(0)}\rangle}{4}
=\frac{1}{4} \sum_{\kk\neq \oo} V_k^{(0)2}+ n_k^{(0)} \left[U_k^{(0)2}+V_k^{(0)2}\right]
\ee
we get as in (\ref{eq:odg_D}) the estimate
\be
\label{eq:estim_FI}
\hat{F}_I \approx (N \epsilon_{\rm Bog})^{1/2}
\ee

\noindent{\bf The hermitian part of $\hat{F}$:} Contrarily to $\hat{F}_I$,
the operator $\hat{F}_R$ alone cannot contribute to the antihermitian part of $\hat{S}_+$, 
it has to be multiplied at least once by the phase difference operator.
To obtain $\hat{S}_y$ up to $\approx N^{1/2}$ included, we thus need $\hat{F}_R$ up to $\approx N$ included.
To this end we decompose $\hat{N}_{\sigma {\bf 0}}$ after the pulse, for $\sigma=a$ or $b$, as follows:
\be
\hat{N}_{\sigma {\bf 0}} = \frac{N}{2} + \hat{\delta {N}}_\sigma \,,  \hspace{0.5cm}  
\hat{\delta N}_\sigma=\left(\hat{N}_\sigma-{N\over 2}\right) -\hat{N}_{\sigma \perp} 
\label{eq:deltaNsigma}
\ee
and we expand the square root in $F$ (\ref{eq:F}) in the large $N$ limit to obtain :
\begin{multline}
\label{eq:devsqrt}
\sqrt{(\hat{N}_{a {\bf 0}}+1) \hat{N}_{b {\bf 0}}} = \frac{N}{2} + \frac{1}{2} (1+\hat{\delta {N}}_a+\hat{\delta {N}}_b) \\
-\frac{(1+\hat{\delta {N}}_a-\hat{\delta {N}}_b)^2}{4N} +\ldots
\end{multline}
In the second contribution in the right-hand side of (\ref{eq:devsqrt}), we can replace $\hat{N}_a+\hat{N}_b$ with
$N$.  Since $\hat{\delta {N}}_a-\hat{\delta {N}}_b$ scales as $N^{1/2}$, the third contribution 
scales as $N^0$ and is thus negligible at the required order. The terms in the $\ldots$
are of too high order in $\epsilon_{\rm size}$ or in $\epsilon_{\rm Bog}$ to be relevant. We conclude that,
for our purposes, we can take
\be
\hat{F}_R \simeq -\frac{1}{2} \sum_\rr dV (\hat{\Lambda}_a^\dagger - \hat{\Lambda}_b^\dagger) (\hat{\Lambda}_a-\hat{\Lambda}_b)
\ee
By expanding the fields $\hat{\Lambda}_a$ and $\hat{\Lambda}_b$ over the Bogoliubov modes, we obtain $\hat{F}_R$ 
in terms of the quasi-particle annihilation operators $\hat{c}_{\sigma \kk}$ at time $t>0$. Using (\ref{eq:ckt})
we can relate these operators to their value at time $0^+$, that we can replace
by the leading order expression (\ref{eq:initczp}) to obtain
\begin{equation}
\hat{F}_R  \simeq - \sum_{\kk \neq 0} \left[ V_k^2 + (U_k^2+V_k^2) \hat{B}_{\kk}^\dagger \hat{B}_{\kk} 
 + U_k V_k \left( e^{-2i \epsilon_k t/\hbar} \hat{B}_{\kk} \hat{B}_{-\kk} 
+ {\rm h.c.} \right) \right]
\label{eq:deltaFR}
\end{equation}
Its expectation value is
\be
\langle \hat{F}_R \rangle = - \sum_{\kk \neq 0} 4 U_k^2 V_k^2 \sin^2\omega_k t
\ee
with $\omega_k=\epsilon_k/\hbar$.
Even if the $\hat{B}$'s correspond to vacuum fluctuations, we still find (replacing the sum by an integral
over $\mathbb{R}^3$, which is convergent,
and making the change of variable $\kk=\KK/\xi_{\rm heal}$, where $\xi_{\rm heal}$ is the healing length)
that $\langle \hat{F}_R \rangle/N$ scales as $(\rho a^3)^{1/2}$, which is $O(\epsilon_{\rm Bog})$.
For simplicity, we shall forget about this detail and consider that $\langle \hat{F}_R \rangle \approx N \epsilon_{\rm Bog}$.
Using Wick's theorem we have also determined the variance of $\hat{F}_R$, 
\be
\mbox{Var}\, \hat{F}_R = \sum_{\kk\neq\oo} 8 U_k^2 V_k^2 \sin^2\omega_k t \left(1+4U_k^2 V_k^2 \sin^2\omega_k t\right)
\ee
Since $(U_k V_k)^4$ diverges as $k^{-4}$ for $k\to 0$, with a similar reasoning as for the discussion around (\ref{eq:var_nperpdiff}), we find that 
$\mbox{Var}\, \hat{F}_R$ is $O\left((N \epsilon_{\rm Bog})^{4/3}\right)$
uniformly in time.  We summarize these estimates by the writing
\be
\hat{F}_R \approx N \epsilon_{\rm Bog} \pm (N \epsilon_{\rm Bog})^{2/3}
\label{eq:estim_FR}
\ee
As we said, we need to estimate $\hat{F}_R$ up to $\approx N$
included. This means that the fluctuations of $\hat{F}_R$, that are
$N^{1/3}$ times smaller, are negligible and the operator $\hat{F}_R$ 
can be replaced by its mean value $\langle \hat{F}_R\rangle$.

\noindent{\bf Operators $\hat{S}_y$, etc:}
From the antihermitian part of (\ref{eq:spd}), and from the estimates  (\ref{eq:estim_difftheta},\ref{eq:estim_FR},
\ref{eq:estim_FI}), we can approximate $\hat{S}_y$ 
up to the terms $\approx N^{1/2}$ included as
\be
\hat{S}_y \simeq \hat{F}_I - \frac{1}{2}\left\{\hat{\theta}_a-\hat{\theta}_b, 
{N\over2}+\langle \hat{F}_R\rangle \right\}\label{eq:SyLT}
\ee
Squaring this expression, and neglecting terms of order larger than one
in $\epsilon_{\rm Bog}$,
we finally obtain the expectation value $\langle \hat{S}_y^2\rangle/N$ 
with an accuracy up to $\approx \epsilon_{\rm size}^0$ 
and $\approx \epsilon_{\rm Bog}$ included, see (\ref{eq:finalsy2}).
Taking the anticommutator of (\ref{eq:SyLT}) with $\hat{S}_z$ and
then the expectation value gives (\ref{eq:finalsysz}), with the additional
simplification that $\langle \hat{F}_I \hat{S}_z\rangle$ is purely imaginary
and cancels out in the anticommutator [this results from
(\ref{eq:namnb}) and (\ref{eq:deltaFI}), and in particular from 
$\langle \hat{y} \hat{x}\rangle =1/(4i)$].

To conclude this Appendix, we give an expectation value useful for subsection \ref{subsec:rotemfx2}:
\be
\langle \{\hat{S}_z,\hat{\cal D}\} \rangle = \sum_{\kk\neq\oo} s_k^2 \left(\langle \hat{A}_\kk^\dagger \hat{A}_\kk\rangle - V_k^2\right)
\label{eq:csls}
\ee
where the $\hat{A}^\dagger \hat{A}$ expectation value is given by (\ref{eq:mada}).
We also give the expression of the difference of non-condensed atom number operators 
at $t>0$:
\begin{multline}
\hat{N}_{a\perp}-\hat{N}_{b\perp} = -\sum_{\kk\neq\mathbf{0}} \Big[(U_k^2+V_k^2) (\hat{A}^\dagger_\kk\hat{B}_\kk+
\hat{B}_\kk^\dagger\hat{A}_\kk)  \\
+2 U_k V_k \left(\hat{A}_\kk\hat{B}_{-\kk} e^{-2i\omega_k t} +\mbox{h.c.}\right)
\Big]
\end{multline}
and the corresponding variance written as a sum of non-negative terms, useful for the discussion below Eq.(\ref{eq:xi02}):
\begin{multline}
\!\!\!\!\mbox{Var}\, (\hat{N}_{a\perp}-\hat{N}_{b\perp}) = \langle N_{a\perp}^{(0)}\rangle 
+\sum_{\kk\neq\mathbf{0}} \frac{1}{2} \left(n_k^{(0)}+\frac{1}{2}\right) \sin^2\omega_kt \\
\!\left[\left(s_k^4-\frac{1}{s_k^{4}}\right)\left(\frac{s_k^4}{s_k^{(0)2}}-\frac{s_k^{(0)2}}{s_k^4}\right) 
+\left(s_k^2-\frac{1}{s_k^{2}}\right)^3\left(\frac{s_k^{(0)2}}{s_k^2}-\frac{s_k^2}{s_k^{(0)2}}\right)\cos^2\omega_kt
\right]
\label{eq:vardnp}
\end{multline}

\section{With the oscillating terms in $\hat{\theta}_a-\hat{\theta}_b$}
\label{app:lourd}

As announced above (\ref{eq:rephase}), we give here the analytical result for $\xi^2(t)$ 
(with the double expansion technique) without performing the approximation used in the main text
of the paper. The temporally oscillating terms in the phase difference operator are now kept,
which amounts to replacing $\hat{\cal D}$ in (\ref{eq:rephase}) with $\hat{\cal D}_{\rm tot}=\hat{\cal D}+\hat{\cal D}_{\rm osc}$
with the oscillating contribution
\be
\hat{\cal D}_{\rm osc}(t) = -\sum_{\kk\neq\oo} s_k^2 \frac{\sin\omega_k t}{\omega_k t} 
\left(e^{-i\omega_k t} \hat{A}_\kk \hat{B}_{-\kk}+\mbox{h.c.}\right)
\ee
where $\omega_k=\epsilon_k/\hbar$.
In the renormalized time (\ref{eq:deftau}) one has also to replace $\hat{\cal D}$ with $\hat{\cal D}_{\rm tot}$,
which involves the new expectation value
\be
\langle \{\hat{S}_z,\hat{\cal D}_{\rm osc} \} \rangle = \sum_{\kk\neq\oo} s_k^2 \frac{\sin 2\omega_k t}{2\omega_k t}
\left(\langle \hat{A}_\kk \hat{A}_{-\kk}\rangle + U_k V_k\right)
\ee
that will thus be added to the contribution (\ref{eq:csls}) [see (\ref{eq:mada}) for the $\hat{A}_\kk^\dagger \hat{A}_{-\kk}$ expectation value].
One can show that $t \langle \{\hat{S}_z,\hat{\cal D}_{\rm osc}\}\rangle$ is uniformly bounded in time, so it contributes
to $\tau$ as a time dependent small temporal shift.
The result (\ref{eq:central}) is replaced by
\be
\xi^2_{\rm tot}(t) \simeq \frac{1-4\frac{\langle \hat{F}_R \rangle}{N}}{(\tau+\sqrt{1+\tau^2})^2} +
\frac{2\left(\frac{\langle\hat{\cal D}_{\rm tot}^2\rangle}{N} \tau^2 + \frac{\langle \hat{F}_R \rangle}{N}+\zeta(t)\right)}
{(\tau+\sqrt{1+\tau^2})\sqrt{1+\tau^2}}
\ee
As expected, again, $\hat{\cal D}$ was replaced by $\hat{\cal D}_{\rm tot}$. There is also an extra term, $\zeta(t)$,
which is simply the value of the last term of (\ref{eq:finalsy2}) at time $t$ minus its value at time $0^+$.
This difference is no longer zero, because (\ref{eq:cslv}) no longer holds when $\hat{\cal D}$
is replaced with $\hat{\cal D}_{\rm tot}$, since $\hat{\cal D}_{\rm tot}$ has imaginary components. We find
\be
N \zeta(t) = -\sum_{\kk\neq\oo} \sin^2\omega_k t \frac{\rho g}{\hbar\omega_k} s_k^2 U_k^{(0)} V_k^{(0)} \left(n_k^{(0)}+\frac{1}{2}\right)
\ee
Note that $\zeta(t)$ is uniformly bounded in time, as $\langle\hat{F}_R\rangle/N$, it is thus not particularly relevant.

More significant deviations may come from the occurrence of $\langle\hat{\cal D}_{\rm tot}^2\rangle$ that differs
from the original $\langle\hat{\cal D}^2\rangle$ by the two terms
\bea
\langle\{\hat{\cal D},\hat{\cal D}_{\rm osc}\}\rangle\!\!\! &=&\!\!\! \sum_{\kk\neq\oo} \! 
\frac{\sin 2\omega_k t}{2\omega_k t}\!
\left(s_k^{(0)2}\!\!-\!\frac{s_k^8}{s_k^{(0)2}}\right)\! (n_k^{(0)}\!\!+\!\frac{1}{2})\\
\langle \hat{\cal D}_{\rm osc}^2\rangle\!\!\! &=&\!\!\! \sum_{\kk\neq\oo} \left(\frac{\sin\omega_k t}{\omega_k t}\right)^2
s_k^4 \Big[(U_k^2+V_k^2)\langle \hat{A}_\kk^\dagger \hat{A}_\kk\rangle \nonumber \\
&& +U_k^2-2U_kV_k\langle\hat{A}_\kk\hat{A}_{-\kk}\rangle
\cos 2\omega_k t\Big]
\eea
These two terms however are $O(\epsilon_{\rm Bog}/\tau^2)$ in the long time limit.
At the relevant times $\tau\approx 1/\epsilon_{\rm Bog}^{1/2}$, where the multimode
nature of the field starts limiting the squeezing, their contributions to $\xi^2_{\rm tot}$ are $O(\epsilon_{\rm Bog}^2)$
and negligible. 

To summarize, the inclusion of the non-oscillating terms in the phase difference operator
does not change at all the long time limit of $\xi^2$ (which, importantly, is its infimum 
in the Bogoliubov approximation): Eq.~(\ref{eq:xi2min_d}) is unchanged. At intermediate times,
it gives small deviations. 
For the extreme case $k_B T/\rho g=10$ and $(\rho a^3)^{1/2}=10^{-3}$, where the non-condensed fraction reaches $10\%$,
we find a maximal relative deviation of $2\%$ between $\xi^2(t)$ of (\ref{eq:central}) and the more accurate
$\xi_{\rm tot}^2(t)$, at a time $\rho g t/\hbar \simeq 1.5$ when  $\xi^2$ is still a factor $\simeq 4$ above
its minimal value $\simeq 0.1$.

\bibliography{paper}
\bibliographystyle{unsrt}

\end{document}